\documentclass[pra,onecolumn,aps,floatfix,amsmath,amssymb,amsfonts]{revtex4-1}
\usepackage{amsmath}
\usepackage{graphicx}
\usepackage{color}

\newcommand{\beq}{\begin{equation}}
\newcommand{\eeq}{\end{equation}}

\graphicspath{{figs_eps/}}

\begin{document}

\title{Reactive collisions in confined geometries}
\author{Zbigniew~Idziaszek$^{1}$, Krzysztof~Jachymski$^{1}$,  and Paul~S.~Julienne$^2$}
\address{$^1$Faculty of Physics, University of Warsaw, Pasteura 5, 02-093 Warsaw, Poland\\
$^2$Joint Quantum Institute, University of Maryland and National Institute of Standards and Technology, College Park, Maryland 20742, USA}

\date{\today}

\begin{abstract}
We consider low energy threshold reactive collisions of particles interacting via a van der Waals potential at long range in the presence of external confinement and give analytic formulas for the confinement modified scattering in such circumstances. The reaction process is described in terms of the short range reaction probability.  Quantum defect theory is used to express elastic and inelastic or reaction collision rates analytically in terms of two dimensionless parameters representing phase and reactivity. We discuss the modifications to Wigner threshold laws for quasi-one-dimensional and quasi-two-dimensional geometries. Confinement-induced resonances are suppressed due to reactions and are completely absent in the universal limit where the short-range loss probability approaches unity.
\end{abstract}

\maketitle
\section{Introduction.} 
Cold and ultracold molecular collisions are an important research topic for a number of reasons, as reviewed by~\cite{Carr2009,PSJ2012}.   In particular, chemical reactions near zero collision energy ($\approx$ 200nK) can be studied experimentally~\cite{Ospelkaus2010,Ni2010} and explained by relatively simple quantum scattering models based on the properties of the long range potential~\cite{Idziaszek2010,Idziaszek2010a}.  Since reactive collisions can result in the rapid loss of trapped molecules, it is important to understand and control them as much as possible.  Reaction rates can be modified and even greatly reduced by aligning dipolar molecules in optical lattice structures of reduced dimensions ~\cite{Quemener2010,Quemener2011,Julienne2011,Zhu2013,Simoni2014}.  

In this work, we present an analytic treatment of ultracold reactive collisions between particles interacting with an isotropic potential, such as S-state atoms or rotationless polar molecules in the absence of an external electric field, confined in a trap that effectively reduces the dimensionality of the system. We parametrize the reaction mechanism at short range using a simple quantum defect parameterization and extend the analytical results  obtained in~\cite{Micheli2010} based on a long range van der Waals potential. In the former work it was assumed that the reaction happens at short range with unit probability, which gives the process several universal features. Here we generalize this treatment to consider the case of non-universal collisions, where the short-range reaction probability is in general smaller than unity. This gives rise to the possibility of resonances in the reaction rates.   Highly reactive molecules in reduced dimensional lattice structures can also experience the Zeno effect, where reaction rates can be suppressed through many-body correlations that develop~\cite{Syassen2008,Durr2009,Zhu2014}.  However, we will not treat such correlations or the Zeno effect here.

A number of workers have developed the idea of using a pseudopotential proportional to the $s$-wave scattering length to represent short range interactions in traps, including highly anisotropic traps that effectively reduce the dimensionality of the system~\cite{Bush,Tiesinga2000,Bolda2002,Blume2002,Bolda2003,Idziaszek2005,Idziaszek2006,Naidon2007,Yurovsky2007,Idziaszek2009}. Quasi-1D systems are of large interest in the context of integrability and exactly solvable models~\cite{Yurovsky2008}. Such trapping potentials can result in confinement-induced resonances (CIR)~\cite{Olshanii,Petrov2000}, which have been studied theoretically~\cite{Petrov2001,Granger2004,Kanjilal2004,Melezhik2009,Drummond2010,Drummond2011,Sala2012} and observed experimentally~\cite{Moritz2005,Haller2010,Sala2013}.  Our theory extends these conventional treatments to give the proper energy-dependent complex scattering length that is needed to calculate such resonances accurately~\cite{Naidon2007}, including also the effect of loss channels due to chemical reactions or inelastic collisions, if they be present. 

This work is structured as follows. In Section 2 we introduce the general scattering problem in the presence of an external trap and the complex scattering length that describes both elastic and inelastic processes.   Section 3 discusses the case of quasi-one-dimensional trap geometry, while Section 4 is dedicated to the quasi-2D case.  In each section we introduce the effective scattering lengths and rate constants and discuss their behavior and possible resonances. Section 5 summarizes the results.

\section{Reactive scattering process in the presence of a trap}
Let us consider two particles confined in an external harmonic trap, which can be described by the stationary Schr\"{o}dinger equation (the center of mass motion has been separated out)
\beq
\label{eq1}
\left[-\frac{\hbar^2}{2\mu}\nabla^2+U(\mathbf{r})+V_{tr}(\mathbf{r})\right]\Psi(\mathbf{r})=E\Psi(\mathbf{r}).
\eeq
Here $\mu$ is the reduced mass of the pair, $U(\mathbf{r})$ is the interparticle interaction and $V_{tr}$ is the harmonic trap, which can confine the particles in two direction $x$ and $y$ (with free ``quasi-1D'' motion along the  $z$ axis) or one direction $z$ (with free ``quasi-2D'' motion in the $x,y$ plane),
\beq
V_{\rm tr\,1D}=\frac{1}{2}\mu \omega^2 \rho^2\quad,\quad V_{\rm tr\,2D}=\frac{1}{2}\mu \omega^2 z^2.
\eeq
It is also possible to consider anisotropic quasi-1D confinement, $ V_{1D}=\frac{1}{2}\mu \omega^2(x^2+\eta y^2)$. These kinds of the trapping potential can be realized in experiment using optical lattices. The harmonic potential is described by characteristic length $d_i=\sqrt{\hbar/\mu\omega_i}$. For each harmonic oscillator state, the total energy $E=E_{iD}$, $i=1,2$,  is composed of the oscillator energy and the free particle energy in the unconfined direction(s)
\beq
E_{1D}=\hbar\omega(1+2n+|m|)+\frac{\hbar^2 p^2}{2\mu}\quad,\quad E_{2D}=\hbar\omega(\nu+\frac{1}{2})+\frac{\hbar^2 q^2}{2\mu}.
\eeq
Here we follow~\cite{Naidon2006}, where the indices $n,m$ denote the state of the 2D harmonic oscillator described by the wave function $\psi_{nm}$, $\nu$ is the state of the 1D harmonic oscillator $\phi_\nu$, and $p$ and $q$ represent the respective quasi-1D and quasi-2D momenta of free motion.
The  asymptotic stationary state solution of  the Schr\"{o}dinger equation is the conventional one representing the sum of an incident plane wave and a scattered wave. The wave function at large distances is then
\begin{equation}
\label{psi1D}
\resizebox{.9\hsize}{!}{$\Psi^{\rm 1D}_{nmp}(\mathbf{r})\stackrel{r\to\infty}{\rightarrow}\psi_{nm}(\mathbf{\rho})e^{ipz}+\sum_{n'm'}{f^{+}_{nm,n'm'}\psi_{n'm'}e^{ip'|z|}}+\sum_{n'm'}{f^{-}_{nm,n'm'}\psi_{n'm'}\frac{z}{|z|}e^{ip'|z|}}$}
\end{equation}
\begin{equation}
\Psi^{\rm 2D}_{\nu q}(\mathbf{r})\stackrel{r\to\infty}{\rightarrow}\phi_\nu(z)e^{i\mathbf{q}\cdot\mathbf{\rho}}+\sum_{\nu'}{f_{\nu\nu'}\phi_{\nu'}(z)\sqrt{\frac{i}{8\pi q'\rho}}e^{i\mathbf{q'}\cdot\mathbf{\rho}}}\,.
\end{equation}

\begin{figure}
\centering
\includegraphics[width=0.5\linewidth]{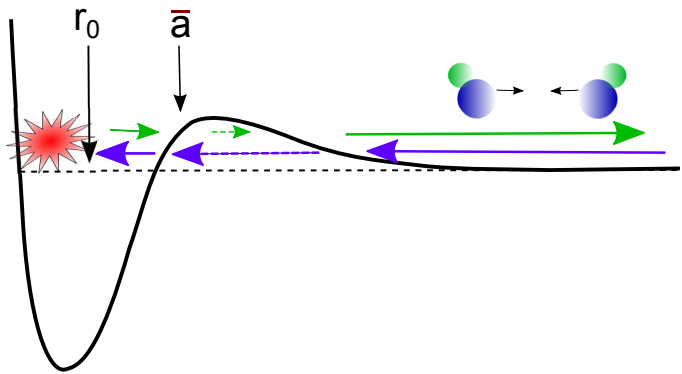}
\caption{\label{interaction}Simple model of a reactive collision. The reaction process takes place at distances $r_0$ much smaller than van der Waals length $\bar{a}$.}
\end{figure}

We will describe the reactive collisions using a simple model based on quantum defect theory. In this treatment the full multichannel interaction potential is replaced by an effective single-channel model with proper boundary conditions at short range~\cite{Idziaszek2010,Jachymski2013}.This single channel represents the $s$- or $p$-wave channel in which the initial ultracold reactant species have been prepared, where $s$ and $p$ respectively represent relative angular momentum quantum numbers $\ell=$ 0 and 1; the former applies to nonidentical species or identical bosons and the latter applies to identical fermions.  The main assumption of the model is the separation of length and energy scales between the chemical reaction and long range scattering processes. The former happens at distances $r_0$ much smaller than typical length scale associated with the long range interactions, which for van der Waals potential $-C_6/r^6$ is given by
\beq
\bar{a}=\frac{2\pi}{\Gamma\left(\frac{1}{4}\right)^2}\left(\frac{2\mu C_6}{\hbar^2}\right)^{1/4}.
\eeq

In our treatment we parametrize the wave function at short range and connect it with the long range solution of the van der Waals potential. The model is schematically presented on Figure~\ref{interaction} and discussed in~\cite{Jachymski2013}. Briefly, if the colliding particles reach the short range, part of the flux will be absorbed there due to reaction or inelastic scattering and part will come back with an additional phase shift. This allows for parametrization of the scattering process using two quantum defect parameters $y$ and $s$, where the $y$ parameter determines the short range reaction probability $P_{\rm re}=\frac{4y}{(1+y)^2}$, and $s$ represents the scattering length of the full interaction potential in units of $\bar{a}$ for scattering in the absence of any loss from the entrance channel.  Essentially, $s$ parameterizes the phase of the wave function due to incoming flux back-scattered into the entrance channel, and $y$ parameterizes any loss of incoming flux due to any inelastic or reactive collision event at short range~\cite{Idziaszek2010}.

In this work we will focus on the case of low energies, where only scattering in the lowest partial wave allowed by the symmetry is relevant. Provided that the van der Waals length scale is much smaller than the confinement length $\bar{a}\ll d$ so that the trapping potential can be regarded as constant  in the interaction range, the scattering process can be described by the $s$ or $p$-wave pseudopotential~\cite{Bolda2002,Bolda2003,Naidon2007,Idziaszek2006,Idziaszek2009}
\begin{eqnarray}
U_s(\mathbf{r})&=\frac{2\pi \hbar^2 a(k)}{\mu}\delta(\mathbf{r})\frac{\partial}{\partial r}r, \label{s-pseudo}\\
U_p(\mathbf{r})&=\frac{\pi \hbar^2 V(k)}{\mu}\stackrel{\leftarrow}{\nabla}\delta(\mathbf{r})\frac{\partial^3}{\partial r^3}r^3 \stackrel{\rightarrow}{\nabla}, \label{p-pseudo}
\end{eqnarray}
where
\begin{eqnarray}
a(k)=-\tan \eta_{\ell=0}(k)/k,\\
V(k)=-\tan \eta_{\ell=1}(k)/k^3
\end{eqnarray}
are the energy-dependent $s$-wave scattering length and $p$-wave scattering volume, respectively and are defined conventionally using the 3D phase shift $\eta_{\ell}(k)$. In the case of reactive collisions these quantities take complex values and can be written in terms of our $y$ and $s$ parameters. For van der Waals interactions~\cite{Jachymski2013}
\begin{eqnarray}
\label{lowka}
a(k)&\stackrel{k\to 0}{\longrightarrow}&\bar{a}\left(s+y\frac{1+(1-s)^2}{y(1-s)+i}\right),\\
\label{lowkv}
V(k)&\stackrel{k\to 0}{\longrightarrow}&-2\bar{V}\frac{y+i(s-1)}{ys+i(s-2)},
\end{eqnarray}
where $\bar{V}=\frac{\pi}{18\Gamma\left(\frac{3}{4}\right)^2}\left(\frac{2\mu C_6}{\hbar^2}\right)^{3/4}$ is the mean $p$-wave scattering volume. 

\section{Quasi-1D case}
It is possible to solve the Schr\"{o}dinger equation~\eqref{eq1} with the boundary condition~\eqref{psi1D} at any energy and find the relation between the scattering amplitudes $f$ and the 3D scattering length for the $s$ wave or volume for the $p$ wave.
At low enough energies $E<2\hbar\omega$ the asymptotic transverse state is the ground state, so we can set the $n,m$ indices to $n=m=0$.   We are then in the quasi-1D regime. Let us first define the quantities relevant for scattering problems in 1D. The 1D $S$ matrix can be connected with the scattering amplitude via $S_{\alpha\alpha}=1+2f_\alpha$~\cite{Naidon2007}, where the ``partial wave'' index $\alpha$ can take on only one of two possible values, corresponding to even ($+$) or odd ($-$) symmetry, and
\beq
f_\alpha(p)=\frac{1}{1+i\cot\eta_\alpha(p)}.
\eeq
Note that the 1D wavenumber $p$ is different from the 3D wavenumber $k$, since $\frac{\hbar^2 k^2}{2\mu}=E=\hbar\omega+\frac{\hbar^2 p^2}{2\mu}$. It is often convenient to use 1D scattering lengths, which can be defined in various ways. Here we choose to define an even and odd scattering ``length'' in the form
\beq
\tilde{a}_\alpha(p)=-p\tan\eta_\alpha(p)=\frac{p}{i}\frac{1-S_{\alpha\alpha}(p)}{1+S_{\alpha\alpha}(p)} \,.
\eeq
Note that the $\tilde{a}_\alpha(p)$ so defined has units of inverse length. Within this notation, the subsequent formulas, including the rate constants, have a form independent of parity. It is also possible to define the 1D scattering lengths which have the unit of length~\cite{Olshanii,Girardeau2004}
\beq
a^{1D}_{\rm e}(p)=\frac{1}{p \tan\eta_{\rm e}(p)} \quad , \quad
a^{1D}_{\rm o}(p)=-\frac{\tan\eta_{\rm o}(p)}{p} .
\eeq
Our quantities can be easily related to these scattering lengths. In the general case $\tilde{a}_\alpha(p)$ will be complex due to the reaction process, $\tilde{a}=\alpha-i\beta$.

From the experimental point of view, the most relevant quantities are the rate constants, in one dimension defined as
\begin{eqnarray}
\label{Kre}
\mathcal{K}_\alpha ^{\rm 1D,\,re}(p)=g\frac{\hbar p}{2\mu}\left(1-|S_{\alpha\alpha}(p)|^2\right),\\
\label{Kel}
\mathcal{K}_\alpha ^{\rm 1D,\,el}(p)=g\frac{\hbar p}{2\mu}\left|1-S_{\alpha\alpha}(p)\right|^2,
\end{eqnarray}
where $g$ is the statistical factor equal to 1 for distinguishable particles and 2 for identical particles in the same internal states. It can be convenient to rewrite these definitions using scattering lengths, obtaining
\begin{eqnarray}
\mathcal{K}_\alpha ^{\rm 1D,\,re}(p)=g\frac{2\hbar p^2}{\mu}\beta_\alpha (p)f^{1D}_\alpha (p)\\
\mathcal{K}_\alpha ^{\rm 1D,\,el}(p)=g\frac{2\hbar p}{\mu}|\tilde{a}_\alpha (p)|^2 f^{1D}_\alpha (p),
\end{eqnarray}
where
\beq
f^{1D}_\alpha (p)=\frac{1}{p^2+|\tilde{a}_\alpha (p)|^2+2p\beta_\alpha (p)}.
\eeq
The loss rate constants determine the decay of one-dimensional particle density $n_{1D}$ of the homogenous gas according to 
\beq
\dot{n}_{1D}=-\mathcal{K}_{\alpha}^{\rm re}n_{1D}^2.
\eeq
Note that $n_{1D}$ has units of (length)$^{-1}$ and $\mathcal{K}_\alpha ^{\rm 1D,\,re}$ has units of (length)$/$(time), so that their product $\mathcal{K}_{\alpha}^{\rm re}n_{1D}$ represents a loss rate per particle that can be compared to the similar 3D loss rate per particle with the conventional 3D rate constant and density. We also note that in the presence of more than two particles, the density decay in one dimension can be greatly affected by  many-body correlations. For example, in the case of strong interactions the particles can form a Tonks-Girardeau gas even in the presence of dissipation, thereby slowing down the reaction rate~\cite{Syassen2008,Durr2009,Zhu2014}.  Treating such a case is beyond the scope of this paper.

\subsection{Even and odd scattering lengths}
Solving the Schr\"{o}dinger equation with the pseudopotentials in Eqs. (\ref{s-pseudo}) - (\ref{p-pseudo}) and boundary conditions~\eqref{psi1D} yields
\begin{eqnarray}
\label{ap}
\frac{1}{\tilde{a}_+(p)}=\frac{d^2}{2a(k)}+\frac{d}{2}\zeta\left(\frac{1}{2}\right),\\
\label{am}
\frac{1}{\tilde{a}_-(p)}=\frac{d^2}{6p^2 V(k)}-\frac{2}{p^2 d}\zeta\left(-\frac{1}{2}\right),
\end{eqnarray}
quite similar to the zero-energy result for elastic scattering. These formulas are only valid in the limit $\frac{\hbar^2 p^2}{2\mu}\ll \hbar \omega$, where $E=\frac{\hbar^2 k^2}{2\mu}\approx\hbar\omega$ so that $k\bar{a}\approx\sqrt{2}\bar{a}/d\ll 1$. The latter condition is also required for the pseudopotential approximation do be valid in the presence of external trap.

Using the low $k$ expansions~\eqref{lowka}-\eqref{lowkv}, the formulas~\eqref{ap}-\eqref{am} can be written as
\begin{eqnarray}
\label{apsy}
\frac{1}{\tilde{a}_+(p)}=\frac{d}{2}\left(\zeta\left(\frac{1}{2}\right)+\frac{d}{\bar{a}}\frac{1+iy(s-1)}{s+iy(s-2)}\right),\\
\label{amsy}
\frac{1}{\tilde{a}_-(p)}=-\frac{2\zeta(-1/2)}{p^2d}-\frac{d^2}{12p^2\bar{V}}\frac{s-2-isy}{s-1-iy}.
\end{eqnarray}
It is instructive to investigate both the $y\to 0$ and $y\to 1$ limits of the expressions above. The former corresponds to the case where reactions are absent and should reduce to the well-known results, while the latter is the universal reactive case for which there should be no dependence on $s$ parameter. Indeed, for $y\to 0$ we recover the formulas of~\cite{Olshanii,Granger2004}:
\begin{eqnarray}
\frac{1}{\tilde{a}_+}\stackrel{y\to 0}{\longrightarrow} \frac{d}{2}\left(\zeta(1/2)+\frac{d}{s\bar{a}}\right),\\
\frac{1}{\tilde{a}_-}\stackrel{y\to 0}{\longrightarrow} -\frac{2\zeta(-1/2)}{p^2d}-\frac{d^2}{12p^2\bar{V}}\frac{s-2}{s-1}.
\end{eqnarray}
In the universal reactive limit
\begin{eqnarray}
\label{aps1}
\frac{1}{\tilde{a}_+(p)}\stackrel{y\to 1}{\longrightarrow}\frac{d}{2}\left(\zeta(1/2)+\frac{d}{\bar{a}(1-i)}\right),\\
\label{ams1}
\frac{1}{\tilde{a}_-(p)}\stackrel{y\to 1}{\longrightarrow}-\frac{2\zeta(-1/2)}{p^2d}-\frac{d^2(1-i)}{12p^2\bar{V}}.
\end{eqnarray}
The 1D scattering lengths define the one-dimensional coupling constants $g_\alpha$ which describe the effective one-dimensional contact interactions $U^{+}_{1D}(z)=g_+\delta(z)$ for even waves and $U^-_{1D}(z)=g_-\stackrel{\leftarrow}{\partial_z}\delta(z)\stackrel{\rightarrow}{\partial_z}$ for odd waves. Using our definitions, we have $g_+(p)=\hbar^2 \tilde{a}_+(p)/\mu$ and $g_-(p)=\hbar^2 \tilde{a}_-(p)/(\mu p^2)$.

\subsection{Rate constants}
We can now calculate the elastic and reactive rate constants using~\eqref{Kre}-\eqref{Kel}. In the limit of very low collision energies $pd^2/\bar{a}\ll 1$, the even scattering rates read
\begin{eqnarray}
\label{Kreelow}
\mathcal{K}_+ ^{\rm 1D,\,re}\stackrel{p\to 0}{\longrightarrow}g\frac{\hbar p^2}{\mu}\frac{d^2 y}{\bar{a}}\frac{2+s(s-2)}{s^2+y^2(s-2)^2},\\
\label{Kelelow}
\mathcal{K}_+ ^{\rm 1D,\,el}\stackrel{p\to 0}{\longrightarrow}g\frac{2\hbar p}{\mu}.
\end{eqnarray}
We note that the elastic rate in the low energy limit approaches a universal value, independent of the reactivity, scattering length and the transverse trap strength.
In the case of odd scattering, the formulas are rather complicated even in the low energy limit $p\bar{V}/d^2\ll 1$, but we provide them for completeness
\begin{eqnarray}
\label{Kreolow}
\mathcal{K}_- ^{\rm 1D,\,re}\stackrel{p\to 0}{\longrightarrow}g\frac{\hbar p^2}{\mu}\frac{12\bar{V}}{d^2}\frac{2y(2+(s-2)s)}{\Xi_{1D}},\\
\label{Kelolow}
\mathcal{K}_- ^{\rm 1D,\,el}\stackrel{p\to 0}{\longrightarrow}g\frac{\hbar p^3}{\mu}\frac{144\bar{V}^2}{d^4}\frac{2(y^2+(1-s)^2)}{\Xi_{1D}},
\end{eqnarray}
where $\Xi_{1D}=4+s(s-4+sy^2)+8\chi+4s\chi(s-3+y^2)+4\chi^2(y^2+(s-1)^2)$ and $\chi=12\bar{V}\zeta(-1/2)/d^3$.
In the universal case $y\to 1$, this yields
\begin{eqnarray}
\mathcal{K}_- ^{\rm 1D,\,re}\stackrel{p\to 0}{\longrightarrow}g\frac{\hbar p^2}{\mu}\frac{12\bar{V}}{d^2}\frac{1}{1+2\chi+2\chi^2},\\
\mathcal{K}_- ^{\rm 1D,\,el}\stackrel{p\to 0}{\longrightarrow}g\frac{\hbar p^3}{2\mu}\frac{144\bar{V}^2}{d^4}\frac{1}{1+2\chi+2\chi^2}.
\end{eqnarray}

\begin{figure}
\centering
\includegraphics[width=0.45\textwidth]{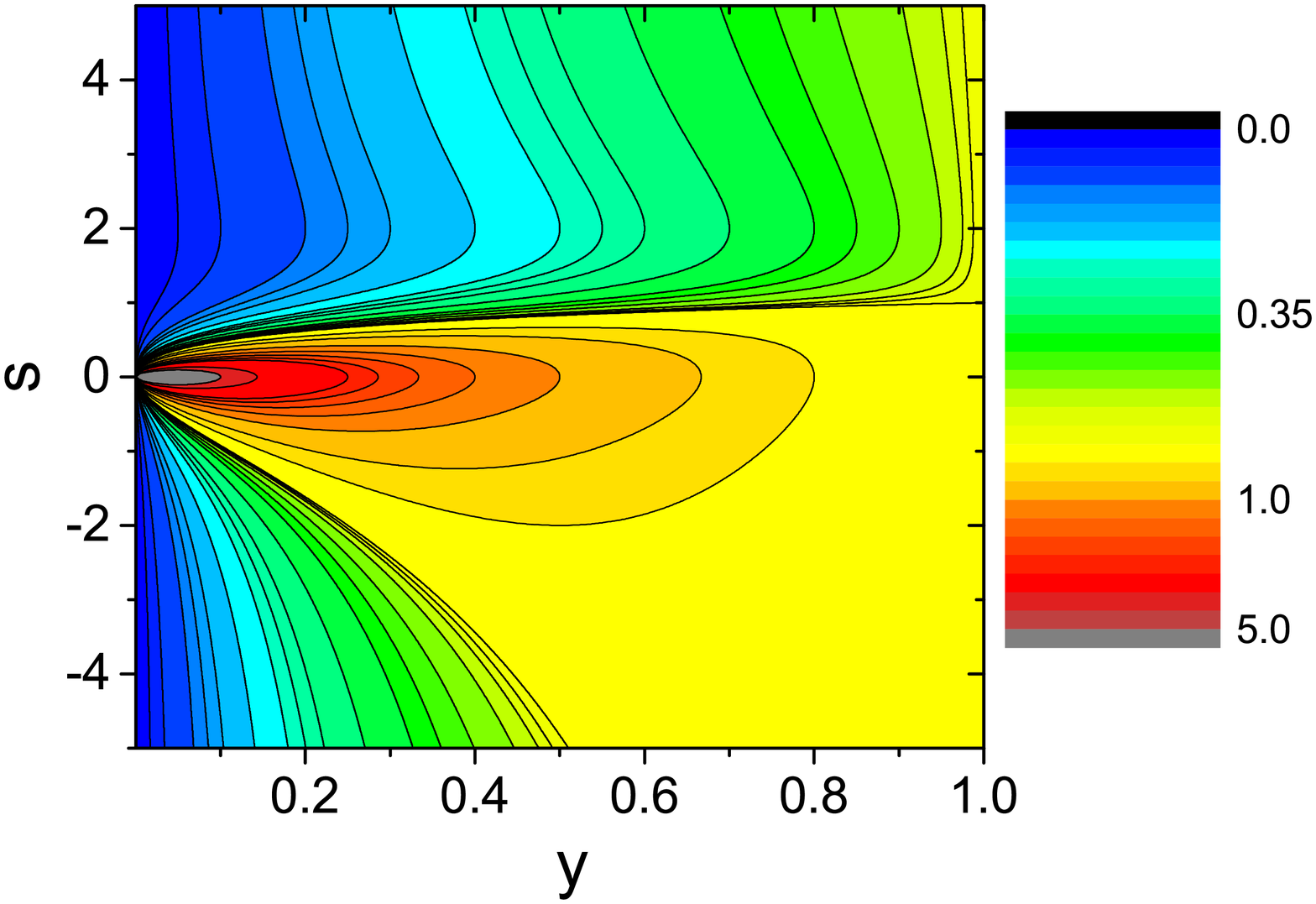}
\includegraphics[width=0.45\textwidth]{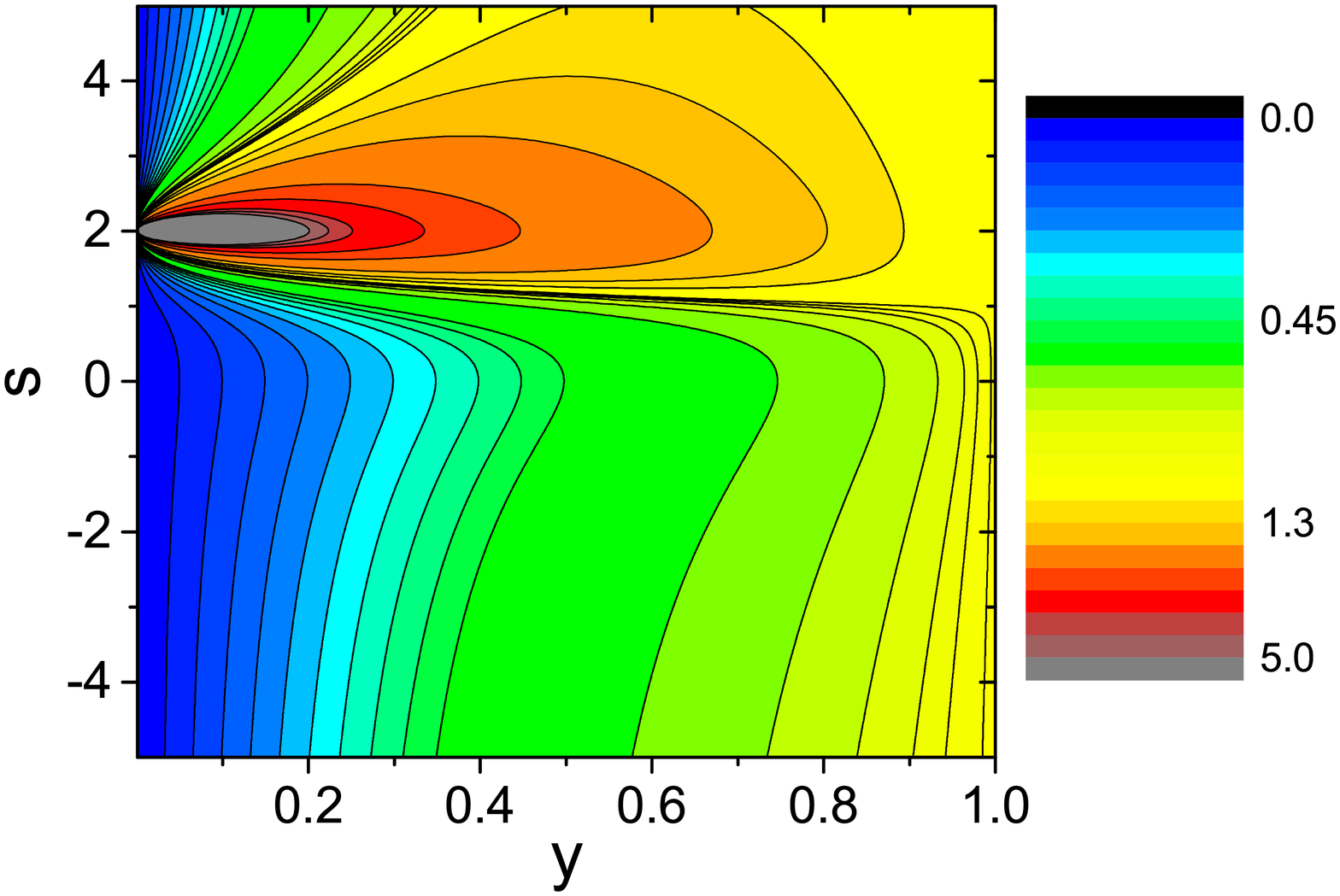}
\caption{\label{k1d}Reactive rate constants rescaled to dimensionless form for quasi-1D even (left) and odd (right) scattering, as given by Eqs.~\eqref{Kreelow} and~\eqref{Kreolow}  (see the main text for details).}
\end{figure}

The low energy behavior of the reactive rate constants is illustrated on Fig.~\ref{k1d}. In both cases only the dimensionless part is plotted for convenience, so in the even case the rate has been divided by $g\hbar p^2 d^2/(\mu\bar{a})$, and in the odd case by $12g\hbar p^2\bar{V}/(\mu d^2)$. For the odd case, in which the denominator still depends on $\bar{V}/d^3$ via the $\chi$ function, a confinement strength corresponding to $d=10\bar{a}$ has been assumed. 
We note that the strongest losses in the even case can be found for low values of $y$ and $s$ close to zero. This stems from the fact that at $p\to 0$ $f^{1D}_+\sim 1/|\tilde{a}_+(p)|^2$. For the odd case, where at low energy $f^{1D}_-\sim 1/p^2$, the reaction rate depends only on the imaginary part of the scattering length. As a result, $s=2$ gives the largest reaction rates. If the confinement is not too strong, $\beta_-$ exhibits a maximum at this particular value associated with a confinement induced resonance, as will be shown in the next section.

\subsection{Impact of reactions on confinement-induced resonance}
\begin{figure}
\centering
\includegraphics[width=0.45\textwidth]{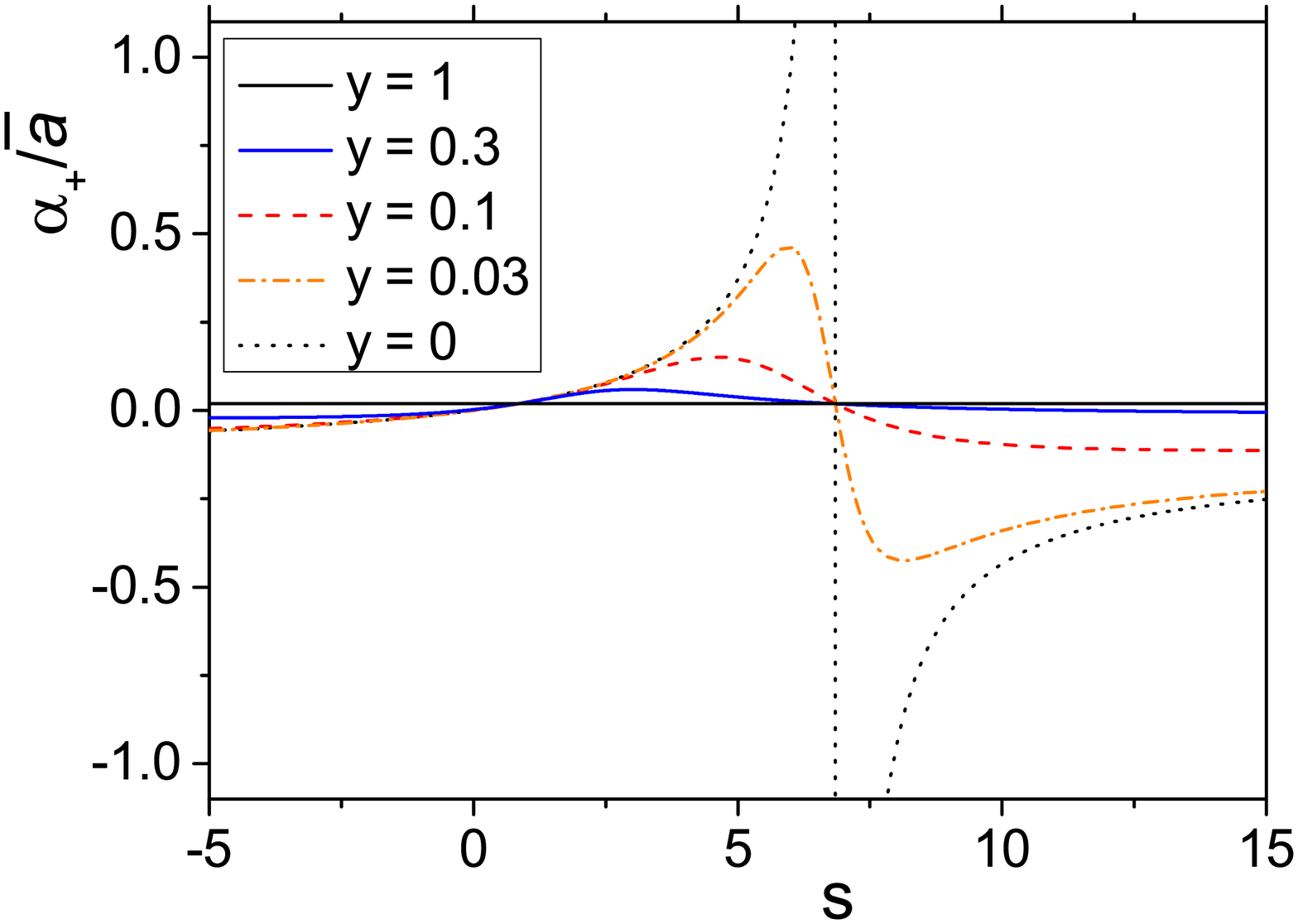}
\includegraphics[width=0.45\textwidth]{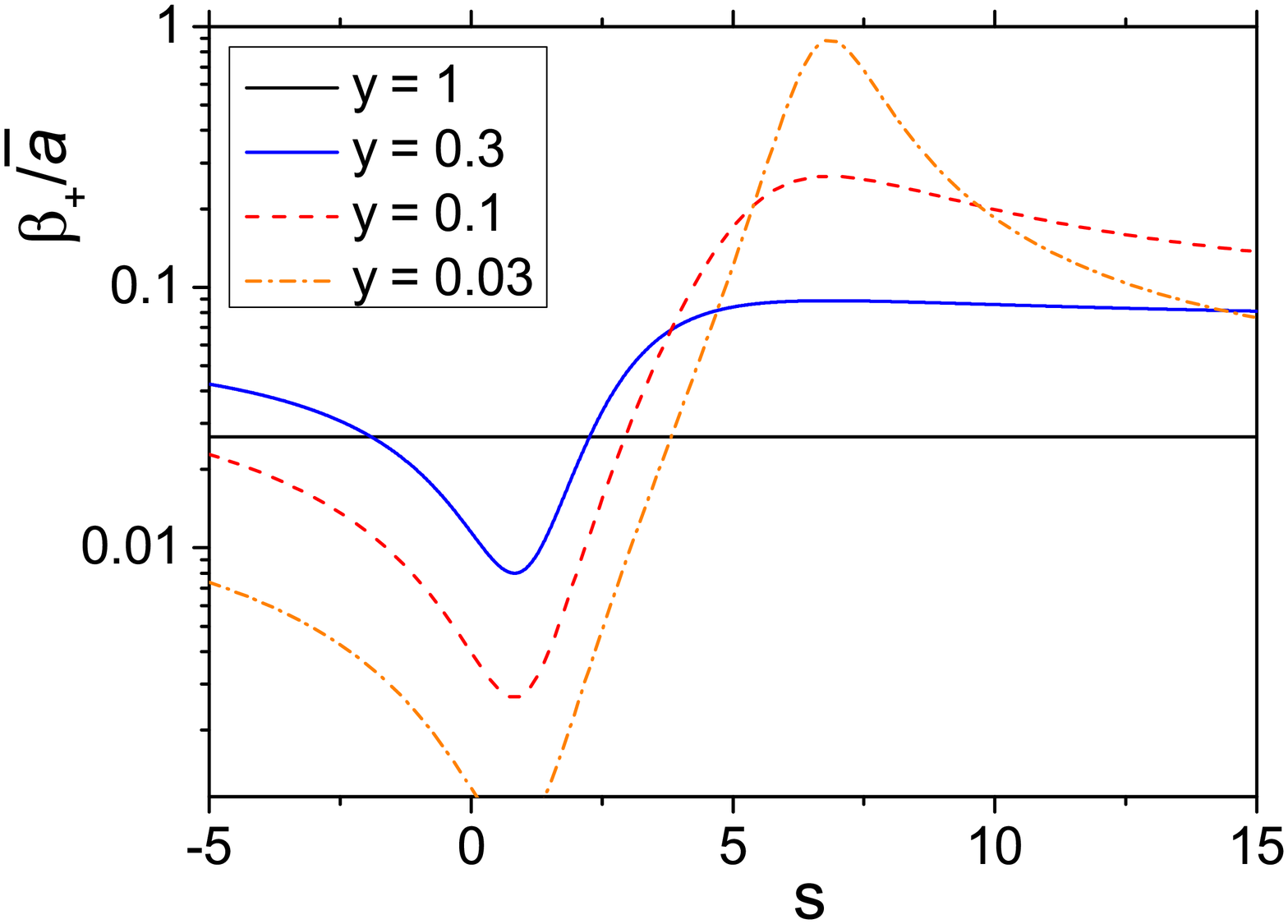}
\caption{\label{a1de}The even one-dimensional scattering length $\tilde{a}_+=\alpha-i\beta$ as a function of the 3D scattering length $s=a_{3D}/\bar{a}$ for different loss parameter $y$. Left: real part $\alpha$, right: imaginary part $\beta$. The transverse confinement corresponds to $d=10\bar{a}$.}
\end{figure}

\begin{figure}
\centering
\includegraphics[width=0.45\textwidth]{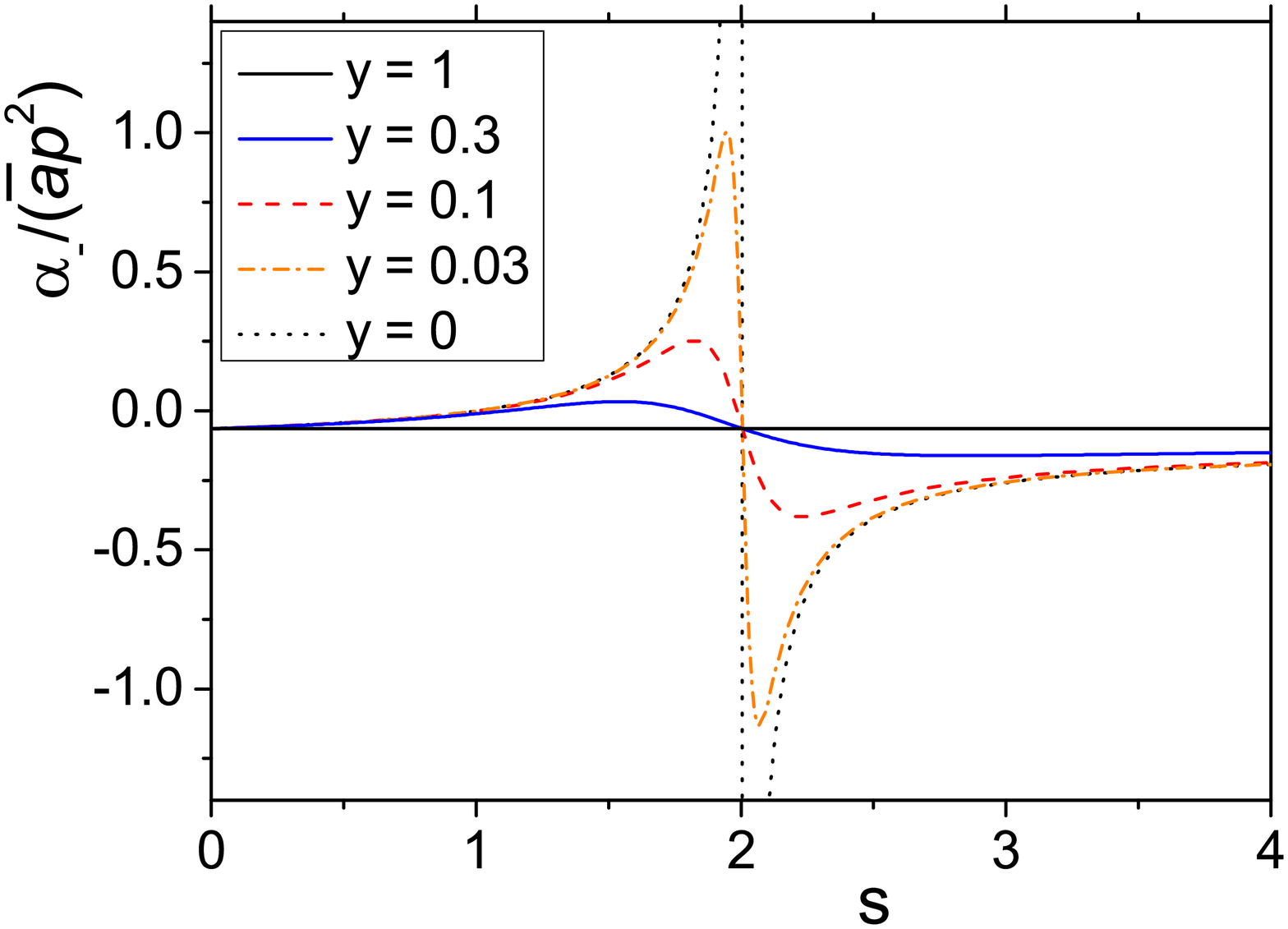}
\includegraphics[width=0.45\textwidth]{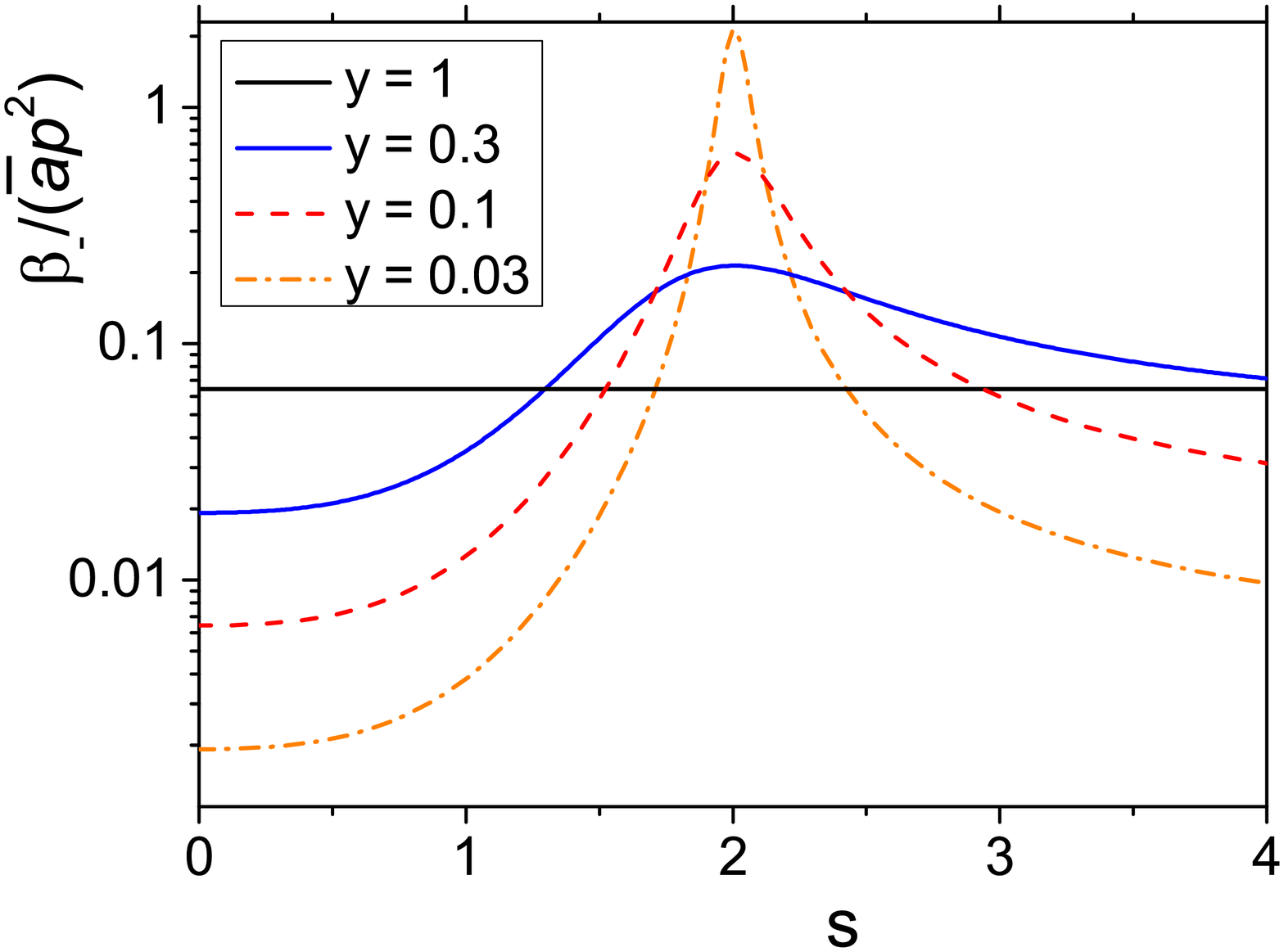}
\caption{\label{a1do} Same as on Fig.~\ref{a1de}, but for the odd scattering length $\tilde{a}_-$. Mind the logarithmic scale used for the imaginary part.}
\end{figure}

The effective 1D coupling constants can diverge if the 3D scattering length is tuned to a certain finite value, which is known as confinement-induced resonance ~\cite{Olshanii}. We will now discuss how the presence of reaction modifies the properties of this resonance. In the absence of reactions, one can calculate at which point the one-dimensional coupling constant approaches infinity and find the resonance at $s^{-1} = -\zeta(1/2)\bar{a}/d$
for even, and  $s=\frac{\chi+1}{\chi+\frac{1}{2}}$ for odd waves, with $\chi$ defined as in Eq.~\eqref{Kreolow}. However, it is easy to verify that the coupling is not divergent if reactions are present. The 1D scattering length, however, is still strongly varying close to the resonance position, and the imaginary part of $\tilde{a}$ exhibits a maximum there. For higher values of the $y$ parameter, this effect gets suppressed, and in the limit of $y\to 1$ the resonance disappears completely. This effect is intuitively clear, since for unit loss probability at short range there is no flux reflected back from short range and thus nothing to ``resonate,'' so no resonances can be present.
The behavior of even and odd scattering lengths at different values of $y$ is illustrated on Figures~\ref{a1de}-\ref{a1do}. We picked $y=0$ (purely elastic collision), $y=0.03$ (very weakly reactive), $y=0.1$ (intermediate case), $y=0.3$ (quite strongly reactive) and $y=1$ (universal reactive) as examples. For $y=0$ we observe the conventional CIR with imaginary part equal to zero, corresponding to no losses. As the reactivity grows, the real parts of the scattering length $\alpha_\pm$  do not diverge anymore and the resonance is washed out. In the universal $y=1$ case $\alpha_\pm$ approaches a constant value. The imaginary parts show quite similar behavior. As soon as the losses are switched on ($y\neq 0$), a sharp peak appears in $\beta_\pm$ at the resonance position. Increasing the value of $y$ makes it less pronounced. In the universal case $\beta_\pm$ does not depend on $s$ anymore. 

We note that it is straightforward to generalize the results from this section to the case of anisotropic transverse confinement, $\omega_x\neq\omega_y$. The resulting formulas are slightly more complicated, but apart from that anisotropy does not introduce any new effects. For example, the even scattering length is given by
\beq
\frac{1}{\tilde{a}_+(p)}=\frac{d_x^2}{2\sqrt{\eta}a_{\ell=0}(k)}+\frac{d C(\eta)}{2\sqrt{\eta}},\\
\eeq
where $\eta=\omega_y/\omega_x$ and $C(\eta)$ is a generalization of $\zeta$ function, defined in~\cite{Drummond2011}.

\section{Quasi-2D case}
We will now analize the case of planar confinement, again assuming that the transverse motion is frozen and we are in the quasi-two-dimensional case. The energy in quasi-2D consists of harmonic oscillator energy and the 2D wavenumber $q$, so in this case  $\frac{\hbar^2 k^2}{2\mu}=E=\frac{1}{2}\hbar\omega+\frac{\hbar^2 q^2}{2\mu}$. For $q\ll 1$ the only relevant term of the 2D scattering amplitude $f_{\nu,\nu'}$ is the $f_{00}$ term, further denoted as $f$. It can be decomposed into ``partial waves'' $f=\sum_{m=-\infty}^\infty{f_m e^{im\Phi}}$. The 2D $S$ matrix is related to the amplitude via $S_{\alpha\alpha}=1+\frac{i}{2}f_{\alpha}$~\cite{Naidon2007} and to the 2D phase shift via
\beq
f_\alpha(q)=\frac{4i}{1+i\cot\eta_\alpha(q)}\,,
\eeq
where the index $\alpha=m$, which is 0 and $\pm 1$ for the two lowest partial waves in this case.
We choose the 2D complex scattering ``length'' to be defined by
\beq
\tilde{a}_\alpha(q)=-\tan\eta_\alpha(q)=\frac{1}{i}\frac{1-S_{\alpha\alpha}(q)}{1+S_{\alpha\alpha}(q)},
\eeq
which is a dimensionless, complex quantity $\tilde{a}_\alpha(q)=\alpha-i\beta$. This choice is again motivated by the simplicity of the following formulas.

\subsection{Scattering lengths}
By solving the Schr\"{o}dinger equation to connect the 2D and 3D scattering lengths, we obtain the following result for the lowest partial waves
\begin{eqnarray}
\frac{1}{\tilde{a}_0(q)}=\frac{1}{\pi}\ln\left(\frac{2B}{\pi q^2 d^2}\right)+\frac{d}{\sqrt{\pi}a(k)},\\
\frac{1}{\tilde{a}_{\pm 1}(q)}=\frac{2d}{3\sqrt{\pi}q^2 V(k)}-\frac{2}{3\pi q^2 d^2}\mathcal{W}(0),
\end{eqnarray}
where $B\approx 0.9049$ and $\mathcal{W}(0)\approx 0.328$~\cite{Idziaszek2006}. Applying formulas~\eqref{lowka}-\eqref{lowkv} to these equations yields
\begin{eqnarray}
\label{a2d0}
\tilde{a}_0(q)=\frac{\sqrt{\pi}\bar{a}}{d}\frac{s(y-i)-2y}{y(s-1)-i+\xi(s(y-i)-2y)},\\
\label{a2d1}
\tilde{a}_{\pm 1}(q)=-\frac{3\sqrt{\pi}\bar{V}q^2}{d}\frac{y+i(s-1)}{\tau(y+i(s-1))+ys+i(s-2)},
\end{eqnarray}
where $\xi=\frac{\bar{a}}{\sqrt{\pi}d}\ln\left(\frac{2B}{\pi q^2 d^2}\right)$ and $\tau=\frac{2\mathcal{W}(0)\bar{V}}{\sqrt{\pi}d^3}$. We note that when
$\bar{a}\ll d$, we have $\tau\ll 1$, but the $\xi$ parameter depends strongly on energy and cannot in general be neglected.
In the universal limit the formulas reduce to
\begin{eqnarray}
\tilde{a}_0(q)\stackrel{y\to 1}{\longrightarrow}\frac{\sqrt{\pi}\bar{a}}{d}\frac{1+2\xi-i}{1+2\xi+2\xi^2},\\
\tilde{a}_{\pm 1}(q)\stackrel{y\to 1}{\longrightarrow}-\frac{3\sqrt{\pi}\bar{V}q^2}{d}\frac{1+\tau+i}{2+2\tau+\tau^2},
\end{eqnarray}
whereas in the nonreactive case
\begin{eqnarray}
\tilde{a}_0(q)\stackrel{y\to 0}{\longrightarrow}\frac{\sqrt{\pi}\bar{a}}{d}\frac{s}{1+s\xi},\\
\tilde{a}_{\pm 1}(q)\stackrel{y\to 0}{\longrightarrow}-\frac{3\sqrt{\pi}\bar{V}q^2}{d}\frac{s-1}{s-2+\tau(s-1)}.
\end{eqnarray}

\subsection{Rate constants}
The two-dimensional reaction rate constants are defined as
\begin{eqnarray}
\label{Kre2d}
\mathcal{K}_\alpha ^{\rm 2D,\,re}(q)=g\frac{\hbar}{\mu}\left(1-|S_{\alpha\alpha}(q)|^2\right),\\
\label{Kel2d}
\mathcal{K}_\alpha ^{\rm 2D,\,el}(q)=g\frac{\hbar}{\mu}\left|1-S_{\alpha\alpha}(q)\right|^2 .
\end{eqnarray}
The decay of two-dimensional density ${n}_{2D}$ is governed by
\beq
\dot{n}_{2D}=-\mathcal{K}_\alpha ^{\rm2D,\, re}n_{2D}^2,
\eeq
where ${n}_{2D}$ has units of (length)$^2$, $\mathcal{K}_\alpha ^{\rm 2D,\,re}$ has units of (length)$^2/$(time), and as for quasi-1D and 3D, their product represents the loss rate per particle.

As in the previous section, it is convenient to rewrite this in the form
\begin{eqnarray}
\mathcal{K}_\alpha ^{\rm 2D,\,re}(q)=g\frac{4\hbar}{\mu}\beta_\alpha (q) f^{2D}_\alpha (q),\\
\mathcal{K}_\alpha ^{\rm 2D,\,el}(q)=g\frac{4\hbar}{\mu}|\tilde{a}_\alpha (q)|^2 f^{2D}_\alpha (q),
\end{eqnarray}
where 
\beq
f^{2D}_\alpha (q)=\frac{1}{1+|\tilde{a}_\alpha (q)|^2+2\beta(q)}.
\eeq

\begin{figure}
\centering
\includegraphics[width=0.45\textwidth]{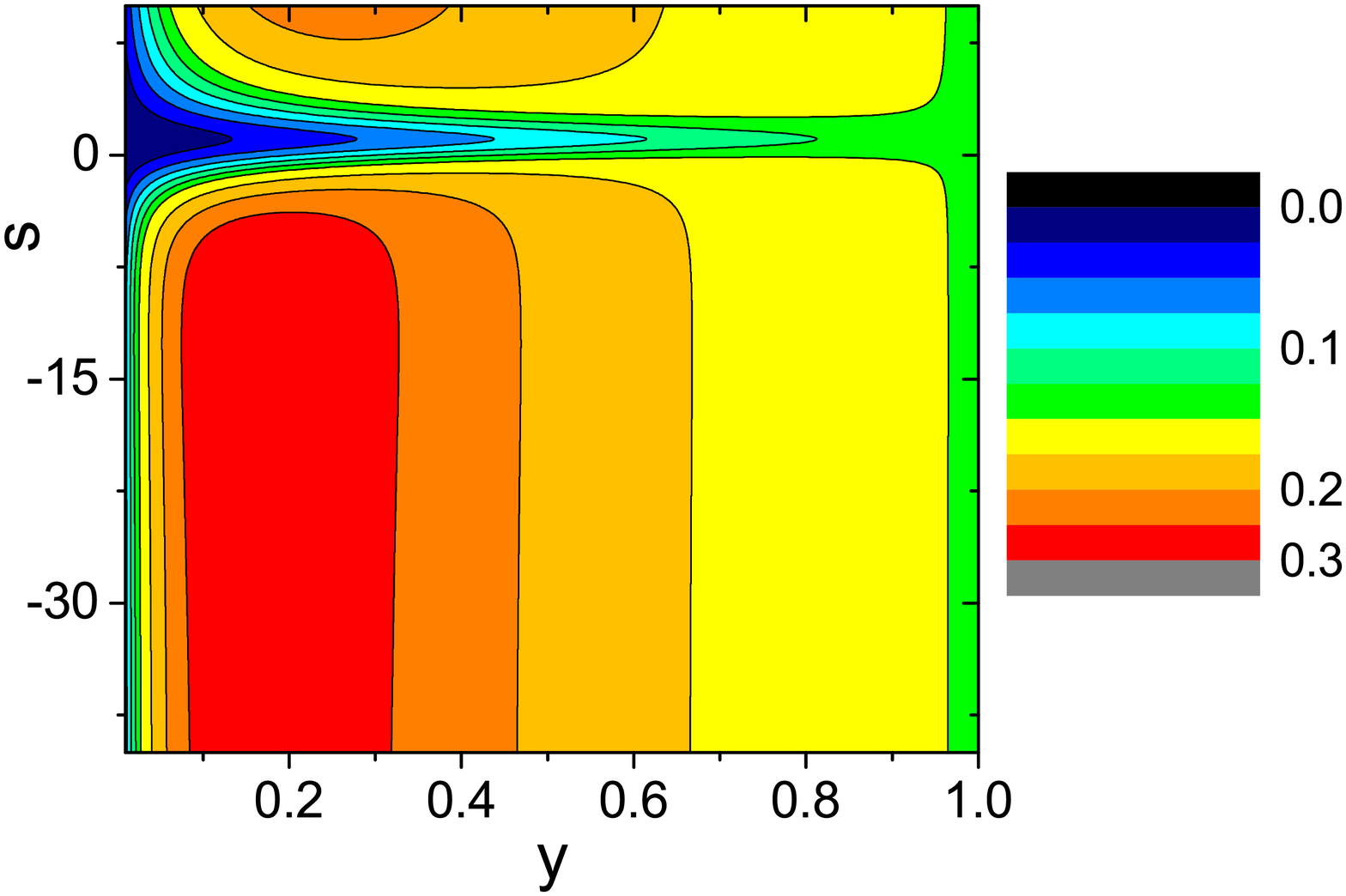}
\includegraphics[width=0.45\textwidth]{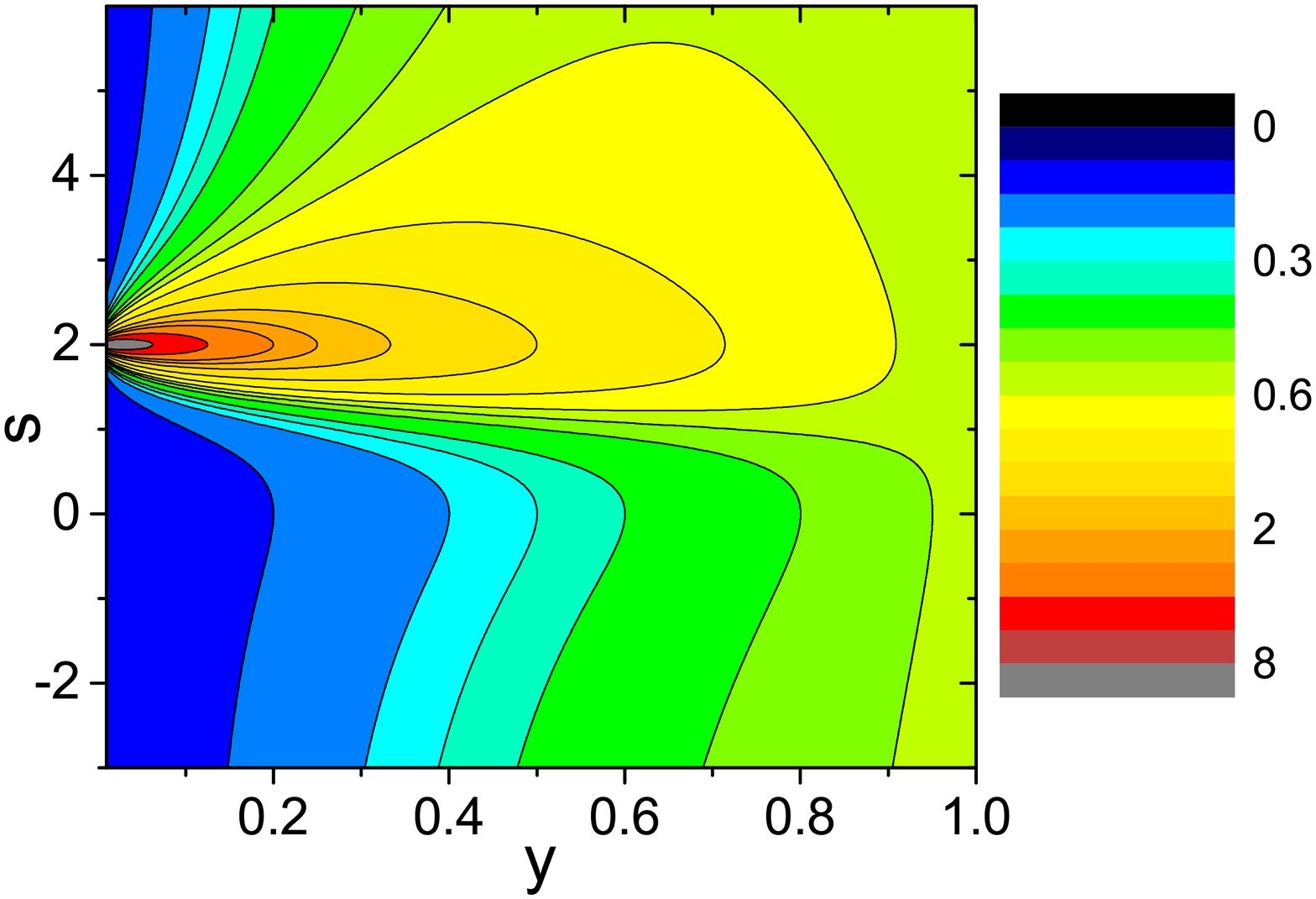}
\caption{\label{k2d}Reactive rate constants rescaled to dimensionless form for quasi-2D $m=0$ (left, given by Eq.~\eqref{Kre2dl0}) and $m=\pm 1$ (right, given by Eq.~\eqref{Kre2dlow}) scattering. The harmonic oscillator length $d=10\bar{a}$ and in the $m=0$ case we assumed $q\bar{a}=0.05$.}
\end{figure}

When plugging in formulas~\eqref{a2d0}-\eqref{a2d1} to obtain the rates, we notice that the rates for $m=0$ have additional energy dependence via the $\xi$ parameter, which contains a term logarithmic in $q^{-2}$. As a result, at $q\to 0$ both rates go to zero logarithmically:
\begin{eqnarray}
\label{Kre2dl0}
\mathcal{K}_{m=0} ^{\rm 2D,\,re}(q)=g\frac{4\hbar\kappa y}{\mu}\frac{s(s-2)+2}{\Xi_{2D,0}}\\
\mathcal{K}_{m=0} ^{\rm 2D,\,el}(q)=g\frac{4\hbar\kappa^2}{\mu}\frac{s^2+(s-2)^2y^2}{\Xi_{2D,0}},
\end{eqnarray}
where $\Xi_{2D,0}=2 \kappa (2+(s-2)s)y+\kappa^2(s^2+(s-2)^2y^2)+y^2(s-1+(s-2)\xi)^2+(1+s\xi)^2$ and $\kappa=\sqrt{\pi}\bar{a}/d$.
In the $m=\pm 1$ case, at sufficiently low energies $q^2 \bar{V}/d\ll 1$, we get
\begin{eqnarray}
\label{Kre2dlow}
\mathcal{K}_{\pm 1} ^{\rm 2D,\,re}(q)\stackrel{q\to 0}{\longrightarrow}g\frac{\hbar}{\mu}\frac{3\sqrt{\pi}\bar{V} q^2}{d}\frac{y(s(s-2)+2)}{\Xi_{2D,\pm1}},\\
\label{Kel2dlow}
\mathcal{K}_{\pm 1} ^{\rm 2D,\,el}(q)\stackrel{q\to 0}{\longrightarrow}g\frac{\hbar}{\mu}\frac{9\pi \bar{V}^2 q^4}{d^2}\frac{y^2+(s-1)^2}{\Xi_{2D,\pm1}},
\end{eqnarray}
where $\Xi_{2D,\pm1}=\tau^2\left(y^2+(s-1)^2\right)+2\tau\left(2-3s+s^2+sy^2\right)+(s-2)^2+s^2y^2$.
Figure \ref{k2d} shows the behavior of the rate constants for different $s$ and $y$. As in the 1D case, only the dimensionless part is plotted. The $m=0$ rate was thus divided by $4g\hbar\kappa/\mu$ and the $m=\pm 1$ rate by $3\sqrt{\pi}g\hbar\bar{V}q^2/(\mu d^2)$.

\subsection{Two-dimensional CIR}
\begin{figure}
\centering
\includegraphics[width=0.45\textwidth]{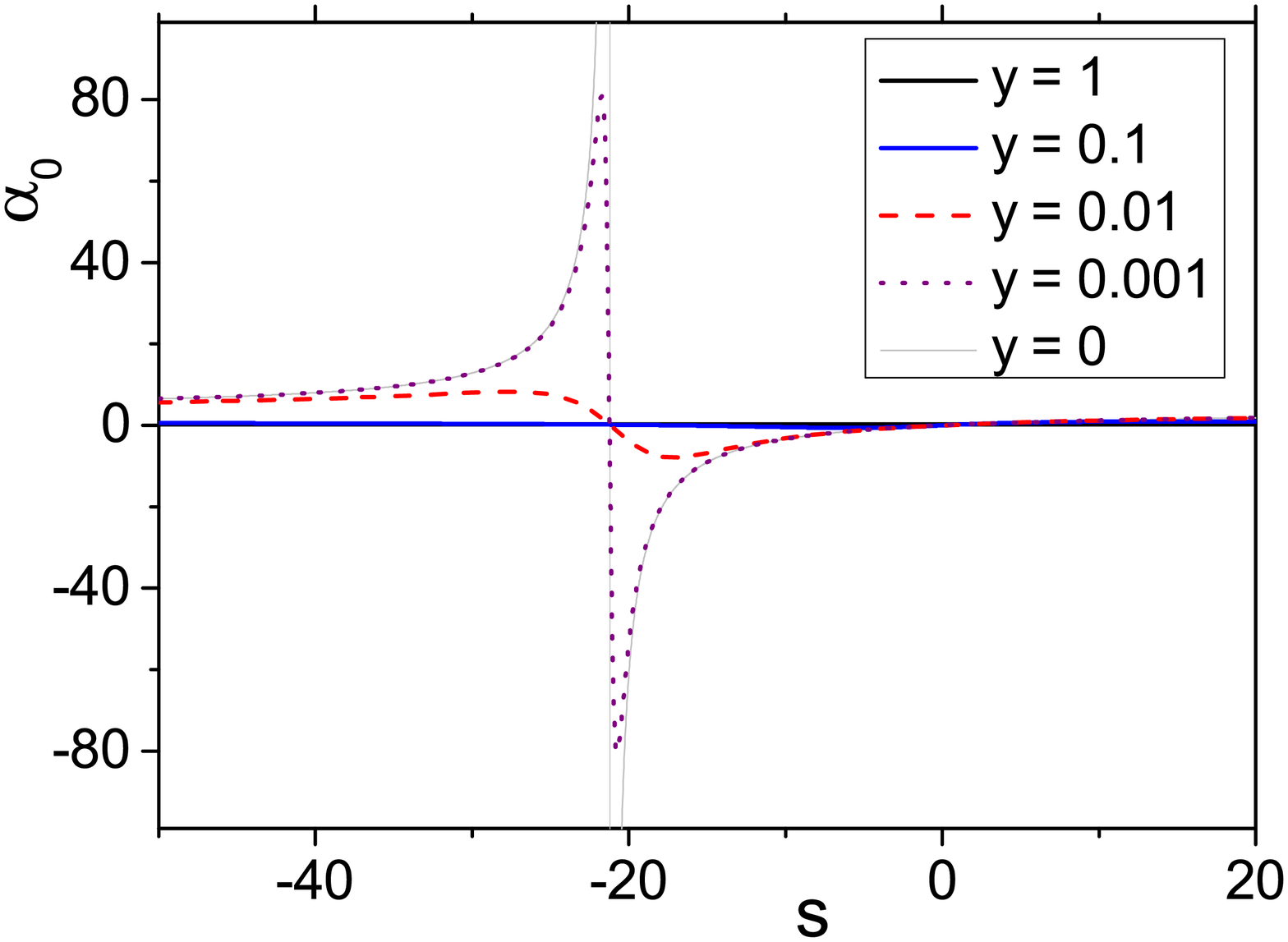}
\includegraphics[width=0.45\textwidth]{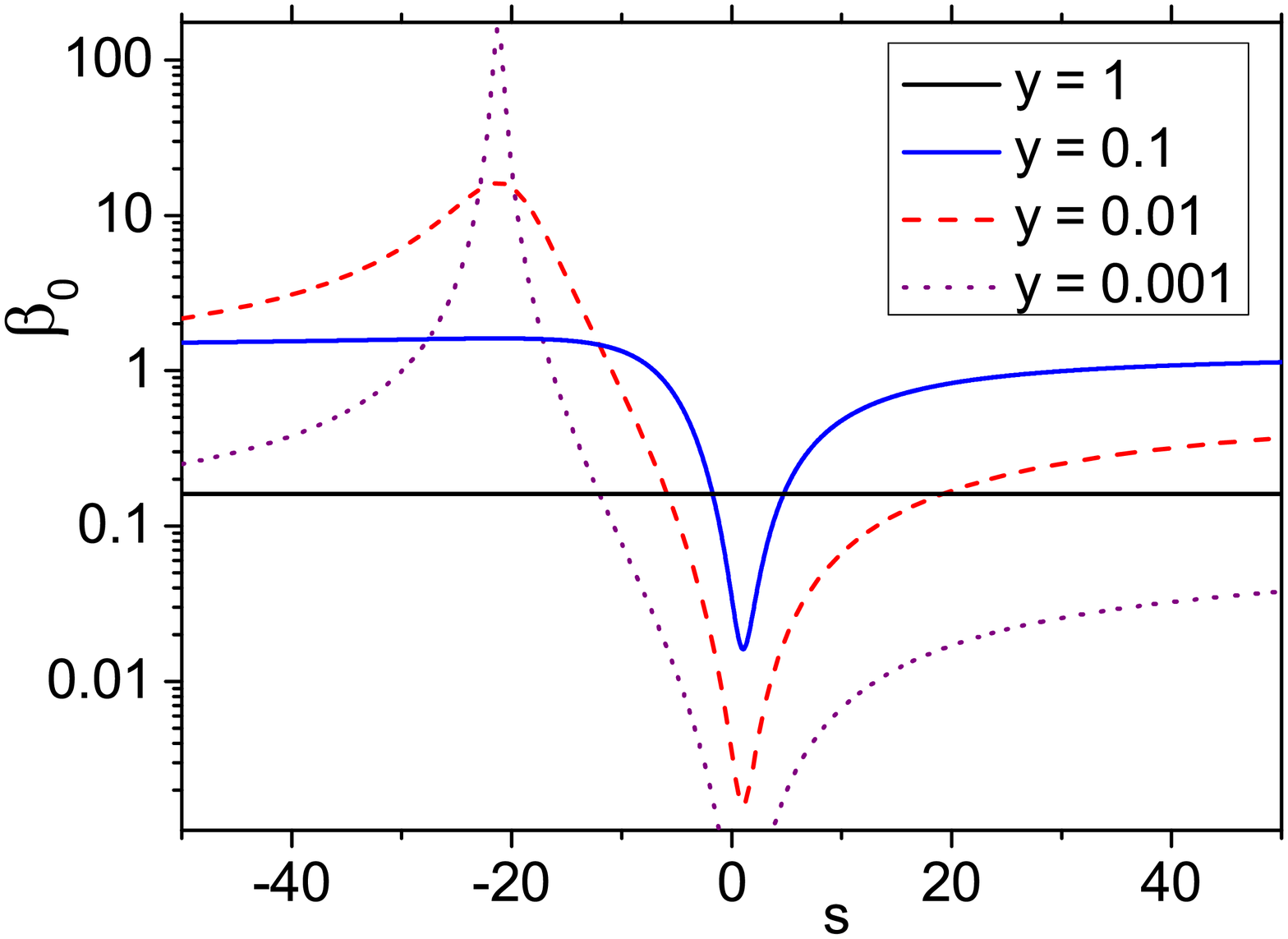}
\caption{\label{a2d0fig}The two-dimensional scattering length $\tilde{a}_0=\alpha-i\beta$ corresponding to $m=0$ as a function of the 3D scattering length $s=a_{3D}/\bar{a}$ for different loss parameters $y$. Left: real part $\alpha$, right: imaginary part $\beta$. The transverse confinement corresponds to $d=10\bar{a}$ and $q\bar{a}=0.05$. Mind the logarithmic scale used for the imaginary part.}
\end{figure}

\begin{figure}
\centering
\includegraphics[width=0.45\textwidth]{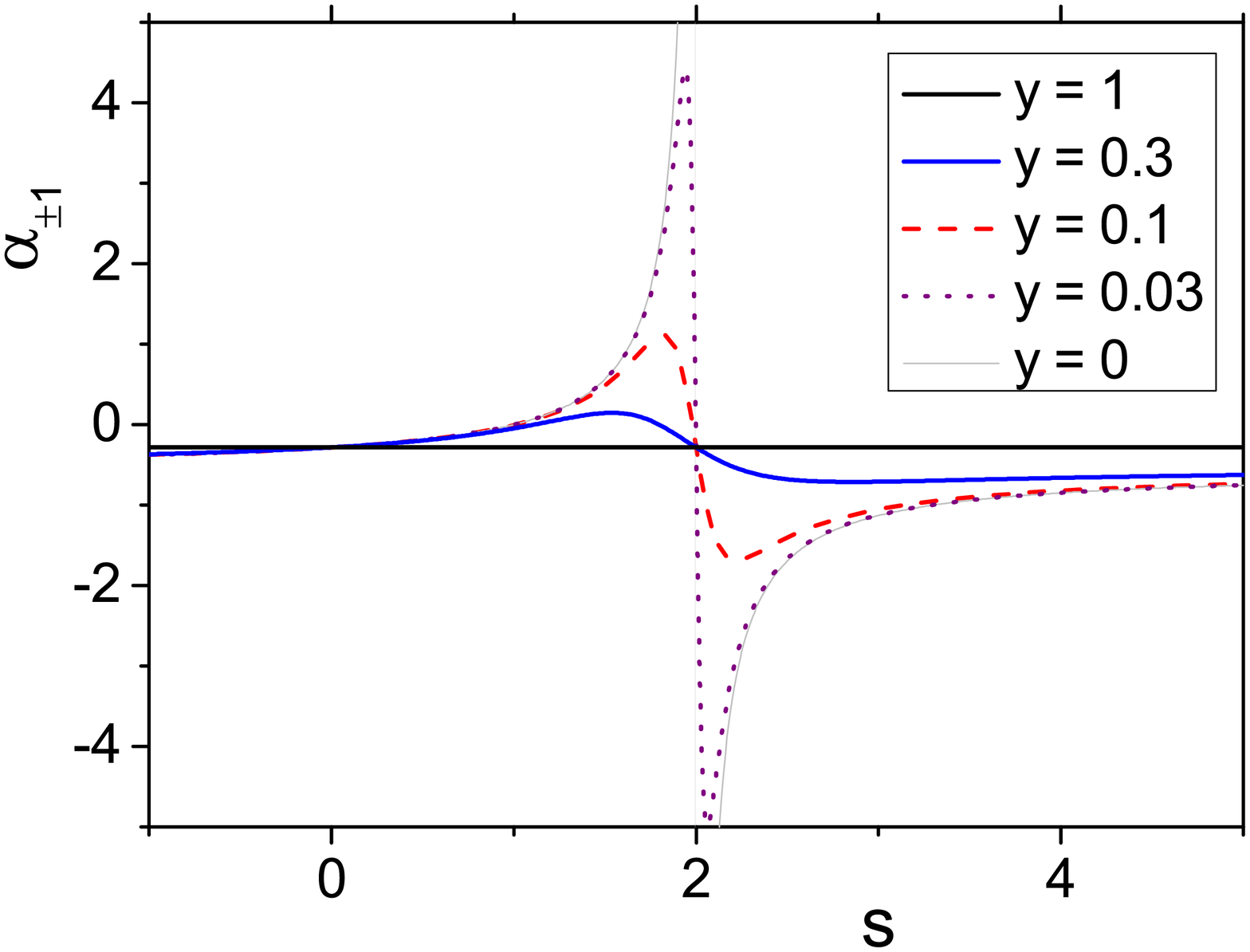}
\includegraphics[width=0.45\textwidth]{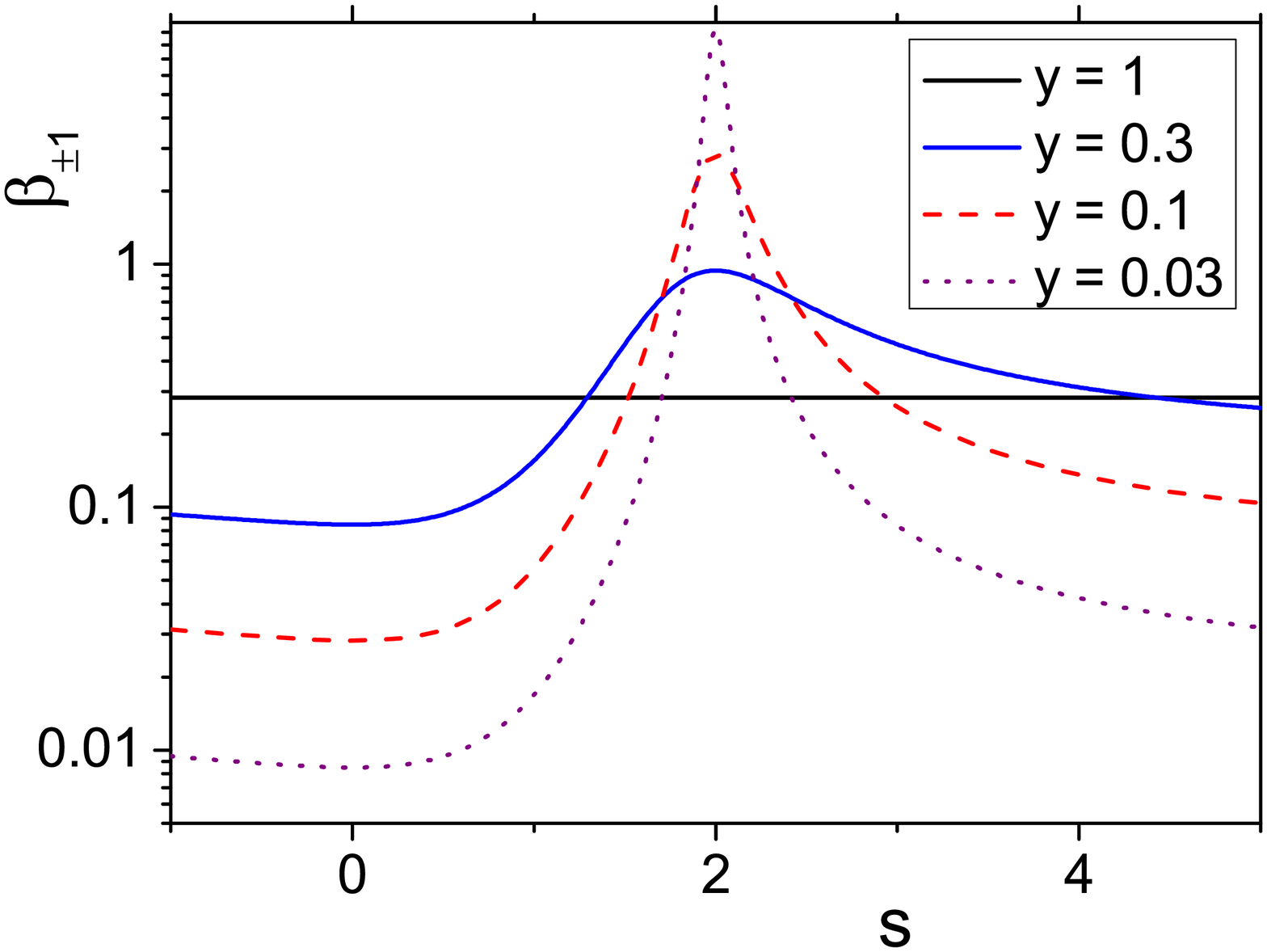}
\caption{\label{a2d1fig} Same as on Fig.~\ref{a2d0fig}, but for $m=\pm 1$ case. Real (left) and imaginary (right) parts of $\tilde{a}_{\pm 1}=\alpha-i\beta$ are shown.}
\end{figure}

In quasi-2D systems, the confinement induced resonance for elastic interactions and $m=0$ can be found by solving the equation $\frac{\sqrt{\pi}d}{s\bar{a}}+\ln\left(\frac{2B}{\pi q^2 d^2}\right)=0$. Its position strongly depends on energy due to the logarithmic term.  In the $m=1$ case, the resonance occurs at $s=\frac{\tau+2}{\tau+1}$ with $\tau$ defined as in~\eqref{a2d1}. Similarly to the quasi-1D case, adding chemical reactions results in finite coupling constant with a pronounced maximum of the imaginary part at the resonance position. Figures~\ref{a2d0fig}-\ref{a2d1fig} show some examples for different values of $y$ from nonreactive to universal case. In the real parts we observe the same kind of behavior as for the quasi-1D case, with a single resonance being washed out as $y\to 1$. For $y=0$ we have $\beta=0$ as before. A sharp resonance appears in the imaginary part for small but nonzero $y$ and is washed out for more reactive collisions. In the $m=0$ case $\beta$ exhibits also a visible minimum at $s$ close to zero. In this regime the real part is also very small. This corresponds to the limit of noninteracting particles. 

\section{Conclusion}

In conclusion, we have given the analytical formulas in the near-threshold limit for the scattering lengths (Eqs.~\eqref{apsy}-\eqref{amsy} nad \eqref{a2d0}-\eqref{a2d1}) as well as elastic and reactive rate constants  (Eqs.~\eqref{Kreelow}-\eqref{Kelolow} nad \eqref{Kre2dl0}-\eqref{Kel2dlow}) for atomic or molecular species interacting by a long range van der Waals potential and undergoing inelastic loss or chemical reactions in the presence of strong confinement of the initial reactant species.  These formulas are based on a powerful quantum defect treatment, which allows the separation of the collision  into a long range and a short range part, with the latter being characterized by two quantum defect parameters.  One, $s$, represents a dimensionless phase and the other, $y$, represents the loss probability of flux from the initially prepared incoming channel of the reactants.  The quantum defect framework  allows the elastic and inelastic or reactive collision rates for quasi-1D or quasi-2D confinement to be expressed in terms of the energy-dependent 3D complex scattering length.  The theory gives analytic predictions with clear intuitive meaning.  While our implementation has been analytic, a numerical implementation is also possible.  This could be useful to extend the theory to species with dipole or quadrupole moments, where more than one inverse power law long range form contributes to the overall potential. It is also possible to generalize the results to a multimode case when several transverse states are occupied.  Our theory also shows how confinement induced resonances are modified in the presence of inelastic or reactive processes.  Such resonances are also potential sources of collision control by manipulating the confinement strength.

This work was supported by the Foundation for Polish Science International PhD Project co-financed by the EU European Regional Development Fund, by National Center for Science grants No. DEC-2011/01/B/ST2/02030 and DEC-2013/09/N/ST2/02188, and by an AFOSR MURI FA9550-09-1-0617.

\section*{References}
\bibliography{allarticles}

\begin{thebibliography}{41}%
\makeatletter
\providecommand \@ifxundefined [1]{%
 \@ifx{#1\undefined}
}%
\providecommand \@ifnum [1]{%
 \ifnum #1\expandafter \@firstoftwo
 \else \expandafter \@secondoftwo
 \fi
}%
\providecommand \@ifx [1]{%
 \ifx #1\expandafter \@firstoftwo
 \else \expandafter \@secondoftwo
 \fi
}%
\providecommand \natexlab [1]{#1}%
\providecommand \enquote  [1]{``#1''}%
\providecommand \bibnamefont  [1]{#1}%
\providecommand \bibfnamefont [1]{#1}%
\providecommand \citenamefont [1]{#1}%
\providecommand \href@noop [0]{\@secondoftwo}%
\providecommand \href [0]{\begingroup \@sanitize@url \@href}%
\providecommand \@href[1]{\@@startlink{#1}\@@href}%
\providecommand \@@href[1]{\endgroup#1\@@endlink}%
\providecommand \@sanitize@url [0]{\catcode `\\12\catcode `\$12\catcode
  `\&12\catcode `\#12\catcode `\^12\catcode `\_12\catcode `\%12\relax}%
\providecommand \@@startlink[1]{}%
\providecommand \@@endlink[0]{}%
\providecommand \url  [0]{\begingroup\@sanitize@url \@url }%
\providecommand \@url [1]{\endgroup\@href {#1}{\urlprefix }}%
\providecommand \urlprefix  [0]{URL }%
\providecommand \Eprint [0]{\href }%
\providecommand \doibase [0]{http://dx.doi.org/}%
\providecommand \selectlanguage [0]{\@gobble}%
\providecommand \bibinfo  [0]{\@secondoftwo}%
\providecommand \bibfield  [0]{\@secondoftwo}%
\providecommand \translation [1]{[#1]}%
\providecommand \BibitemOpen [0]{}%
\providecommand \bibitemStop [0]{}%
\providecommand \bibitemNoStop [0]{.\EOS\space}%
\providecommand \EOS [0]{\spacefactor3000\relax}%
\providecommand \BibitemShut  [1]{\csname bibitem#1\endcsname}%
\let\auto@bib@innerbib\@empty
\bibitem [{\citenamefont {Carr}\ \emph {et~al.}(2009)\citenamefont {Carr},
  \citenamefont {DeMille}, \citenamefont {Krems},\ and\ \citenamefont
  {Ye}}]{Carr2009}%
  \BibitemOpen
  \bibfield  {author} {\bibinfo {author} {\bibfnamefont {L.~D.}\ \bibnamefont
  {Carr}}, \bibinfo {author} {\bibfnamefont {D.}~\bibnamefont {DeMille}},
  \bibinfo {author} {\bibfnamefont {R.~V.}\ \bibnamefont {Krems}}, \ and\
  \bibinfo {author} {\bibfnamefont {J.}~\bibnamefont {Ye}},\ }\href@noop {}
  {\bibfield  {journal} {\bibinfo  {journal} {New Journ. Phys.}\ }\textbf
  {\bibinfo {volume} {11}},\ \bibinfo {pages} {055049} (\bibinfo {year}
  {2009})}\BibitemShut {NoStop}%
\bibitem [{\citenamefont {Qu\`{e}m\`{e}ner}\ and\ \citenamefont
  {Julienne}(2012)}]{PSJ2012}%
  \BibitemOpen
  \bibfield  {author} {\bibinfo {author} {\bibfnamefont {G.}~\bibnamefont
  {Qu\`{e}m\`{e}ner}}\ and\ \bibinfo {author} {\bibfnamefont {P.~S.}\
  \bibnamefont {Julienne}},\ }\href@noop {} {\bibfield  {journal} {\bibinfo
  {journal} {Chemical Reviews}\ }\textbf {\bibinfo {volume} {112}},\ \bibinfo
  {pages} {4949} (\bibinfo {year} {2012})}\BibitemShut {NoStop}%
\bibitem [{\citenamefont {Ospelkaus}\ \emph {et~al.}(2010)\citenamefont
  {Ospelkaus}, \citenamefont {Ni}, \citenamefont {Wang}, \citenamefont
  {de~Miranda}, \citenamefont {Neyenhuis}, \citenamefont {Qu\'{e}m\'{e}ner},
  \citenamefont {Julienne}, \citenamefont {Bohn}, \citenamefont {Jin},\ and\
  \citenamefont {Ye}}]{Ospelkaus2010}%
  \BibitemOpen
  \bibfield  {author} {\bibinfo {author} {\bibfnamefont {S.}~\bibnamefont
  {Ospelkaus}}, \bibinfo {author} {\bibfnamefont {K.-K.}\ \bibnamefont {Ni}},
  \bibinfo {author} {\bibfnamefont {D.}~\bibnamefont {Wang}}, \bibinfo {author}
  {\bibfnamefont {M.~H.~G.}\ \bibnamefont {de~Miranda}}, \bibinfo {author}
  {\bibfnamefont {B.}~\bibnamefont {Neyenhuis}}, \bibinfo {author}
  {\bibfnamefont {G.}~\bibnamefont {Qu\'{e}m\'{e}ner}}, \bibinfo {author}
  {\bibfnamefont {P.~S.}\ \bibnamefont {Julienne}}, \bibinfo {author}
  {\bibfnamefont {J.~L.}\ \bibnamefont {Bohn}}, \bibinfo {author}
  {\bibfnamefont {D.~S.}\ \bibnamefont {Jin}}, \ and\ \bibinfo {author}
  {\bibfnamefont {J.}~\bibnamefont {Ye}},\ }\href@noop {} {\bibfield  {journal}
  {\bibinfo  {journal} {Science}\ }\textbf {\bibinfo {volume} {327}},\ \bibinfo
  {pages} {853} (\bibinfo {year} {2010})}\BibitemShut {NoStop}%
\bibitem [{\citenamefont {Ni}\ \emph {et~al.}(2010)\citenamefont {Ni},
  \citenamefont {Ospelkaus}, \citenamefont {Wang}, \citenamefont
  {Qu{\'e}m{\'e}ner}, \citenamefont {Neyenhuis}, \citenamefont {de~Miranda},
  \citenamefont {Bohn}, \citenamefont {Ye},\ and\ \citenamefont
  {Jin}}]{Ni2010}%
  \BibitemOpen
  \bibfield  {author} {\bibinfo {author} {\bibfnamefont {K.-K.}\ \bibnamefont
  {Ni}}, \bibinfo {author} {\bibfnamefont {S.}~\bibnamefont {Ospelkaus}},
  \bibinfo {author} {\bibfnamefont {D.}~\bibnamefont {Wang}}, \bibinfo {author}
  {\bibfnamefont {G.}~\bibnamefont {Qu{\'e}m{\'e}ner}}, \bibinfo {author}
  {\bibfnamefont {B.}~\bibnamefont {Neyenhuis}}, \bibinfo {author}
  {\bibfnamefont {M.~H.~G.}\ \bibnamefont {de~Miranda}}, \bibinfo {author}
  {\bibfnamefont {J.~L.}\ \bibnamefont {Bohn}}, \bibinfo {author}
  {\bibfnamefont {J.}~\bibnamefont {Ye}}, \ and\ \bibinfo {author}
  {\bibfnamefont {D.~S.}\ \bibnamefont {Jin}},\ }\href@noop {} {\bibfield
  {journal} {\bibinfo  {journal} {Nature}\ }\textbf {\bibinfo {volume} {464}},\
  \bibinfo {pages} {1324} (\bibinfo {year} {2010})}\BibitemShut {NoStop}%
\bibitem [{\citenamefont {Idziaszek}\ and\ \citenamefont
  {Julienne}(2010)}]{Idziaszek2010}%
  \BibitemOpen
  \bibfield  {author} {\bibinfo {author} {\bibfnamefont {Z.}~\bibnamefont
  {Idziaszek}}\ and\ \bibinfo {author} {\bibfnamefont {P.~S.}\ \bibnamefont
  {Julienne}},\ }\href@noop {} {\bibfield  {journal} {\bibinfo  {journal}
  {Phys. Rev. Lett.}\ }\textbf {\bibinfo {volume} {104}},\ \bibinfo {pages}
  {113202} (\bibinfo {year} {2010})}\BibitemShut {NoStop}%
\bibitem [{\citenamefont {Idziaszek}\ \emph {et~al.}(2010)\citenamefont
  {Idziaszek}, \citenamefont {Qu\'em\'ener}, \citenamefont {Bohn},\ and\
  \citenamefont {Julienne}}]{Idziaszek2010a}%
  \BibitemOpen
  \bibfield  {author} {\bibinfo {author} {\bibfnamefont {Z.}~\bibnamefont
  {Idziaszek}}, \bibinfo {author} {\bibfnamefont {G.}~\bibnamefont
  {Qu\'em\'ener}}, \bibinfo {author} {\bibfnamefont {J.~L.}\ \bibnamefont
  {Bohn}}, \ and\ \bibinfo {author} {\bibfnamefont {P.~S.}\ \bibnamefont
  {Julienne}},\ }\href@noop {} {\bibfield  {journal} {\bibinfo  {journal}
  {Phys. Rev. A}\ }\textbf {\bibinfo {volume} {82}},\ \bibinfo {pages} {020703}
  (\bibinfo {year} {2010})}\BibitemShut {NoStop}%
\bibitem [{\citenamefont {Qu\'em\'ener}\ and\ \citenamefont
  {Bohn}(2010)}]{Quemener2010}%
  \BibitemOpen
  \bibfield  {author} {\bibinfo {author} {\bibfnamefont {G.}~\bibnamefont
  {Qu\'em\'ener}}\ and\ \bibinfo {author} {\bibfnamefont {J.~L.}\ \bibnamefont
  {Bohn}},\ }\href@noop {} {\bibfield  {journal} {\bibinfo  {journal} {Phys.
  Rev. A}\ }\textbf {\bibinfo {volume} {81}},\ \bibinfo {pages} {060701}
  (\bibinfo {year} {2010})}\BibitemShut {NoStop}%
\bibitem [{\citenamefont {Qu\'em\'ener}\ and\ \citenamefont
  {Bohn}(2011)}]{Quemener2011}%
  \BibitemOpen
  \bibfield  {author} {\bibinfo {author} {\bibfnamefont {G.}~\bibnamefont
  {Qu\'em\'ener}}\ and\ \bibinfo {author} {\bibfnamefont {J.~L.}\ \bibnamefont
  {Bohn}},\ }\href@noop {} {\bibfield  {journal} {\bibinfo  {journal} {Phys.
  Rev. A}\ }\textbf {\bibinfo {volume} {83}},\ \bibinfo {pages} {012705}
  (\bibinfo {year} {2011})}\BibitemShut {NoStop}%
\bibitem [{\citenamefont {Julienne}\ \emph {et~al.}(2011)\citenamefont
  {Julienne}, \citenamefont {Hanna},\ and\ \citenamefont
  {Idziaszek}}]{Julienne2011}%
  \BibitemOpen
  \bibfield  {author} {\bibinfo {author} {\bibfnamefont {P.~S.}\ \bibnamefont
  {Julienne}}, \bibinfo {author} {\bibfnamefont {T.~M.}\ \bibnamefont {Hanna}},
  \ and\ \bibinfo {author} {\bibfnamefont {Z.}~\bibnamefont {Idziaszek}},\
  }\href@noop {} {\bibfield  {journal} {\bibinfo  {journal} {Physical Chemistry
  Chemical Physics}\ }\textbf {\bibinfo {volume} {13}},\ \bibinfo {pages}
  {19114} (\bibinfo {year} {2011})}\BibitemShut {NoStop}%
\bibitem [{\citenamefont {Zhu}\ \emph {et~al.}(2013)\citenamefont {Zhu},
  \citenamefont {Qu\'em\'ener}, \citenamefont {Rey},\ and\ \citenamefont
  {Holland}}]{Zhu2013}%
  \BibitemOpen
  \bibfield  {author} {\bibinfo {author} {\bibfnamefont {B.}~\bibnamefont
  {Zhu}}, \bibinfo {author} {\bibfnamefont {G.}~\bibnamefont {Qu\'em\'ener}},
  \bibinfo {author} {\bibfnamefont {A.~M.}\ \bibnamefont {Rey}}, \ and\
  \bibinfo {author} {\bibfnamefont {M.~J.}\ \bibnamefont {Holland}},\
  }\href@noop {} {\bibfield  {journal} {\bibinfo  {journal} {Phys. Rev. A}\
  }\textbf {\bibinfo {volume} {88}},\ \bibinfo {pages} {063405} (\bibinfo
  {year} {2013})}\BibitemShut {NoStop}%
\bibitem [{\citenamefont {Simoni}\ \emph {et~al.}(2015)\citenamefont {Simoni},
  \citenamefont {Srinivasan}, \citenamefont {Launay}, \citenamefont
  {Jachymski}, \citenamefont {Idziaszek},\ and\ \citenamefont
  {Julienne}}]{Simoni2014}%
  \BibitemOpen
  \bibfield  {author} {\bibinfo {author} {\bibfnamefont {A.}~\bibnamefont
  {Simoni}}, \bibinfo {author} {\bibfnamefont {S.}~\bibnamefont {Srinivasan}},
  \bibinfo {author} {\bibfnamefont {J.-M.}\ \bibnamefont {Launay}}, \bibinfo
  {author} {\bibfnamefont {K.}~\bibnamefont {Jachymski}}, \bibinfo {author}
  {\bibfnamefont {Z.}~\bibnamefont {Idziaszek}}, \ and\ \bibinfo {author}
  {\bibfnamefont {P.~S.}\ \bibnamefont {Julienne}},\ }\href@noop {} {\bibfield
  {journal} {\bibinfo  {journal} {New Journal of Physics}\ }\textbf {\bibinfo
  {volume} {17}},\ \bibinfo {pages} {013020} (\bibinfo {year}
  {2015})}\BibitemShut {NoStop}%
\bibitem [{\citenamefont {Micheli}\ \emph {et~al.}(2010)\citenamefont
  {Micheli}, \citenamefont {Idziaszek}, \citenamefont {Pupillo}, \citenamefont
  {Baranov}, \citenamefont {Zoller},\ and\ \citenamefont
  {Julienne}}]{Micheli2010}%
  \BibitemOpen
  \bibfield  {author} {\bibinfo {author} {\bibfnamefont {A.}~\bibnamefont
  {Micheli}}, \bibinfo {author} {\bibfnamefont {Z.}~\bibnamefont {Idziaszek}},
  \bibinfo {author} {\bibfnamefont {G.}~\bibnamefont {Pupillo}}, \bibinfo
  {author} {\bibfnamefont {M.~A.}\ \bibnamefont {Baranov}}, \bibinfo {author}
  {\bibfnamefont {P.}~\bibnamefont {Zoller}}, \ and\ \bibinfo {author}
  {\bibfnamefont {P.~S.}\ \bibnamefont {Julienne}},\ }\href@noop {} {\bibfield
  {journal} {\bibinfo  {journal} {Phys. Rev. Lett.}\ }\textbf {\bibinfo
  {volume} {105}},\ \bibinfo {pages} {073202} (\bibinfo {year}
  {2010})}\BibitemShut {NoStop}%
\bibitem [{\citenamefont {Syassen}\ \emph {et~al.}(2008)\citenamefont
  {Syassen}, \citenamefont {Bauer}, \citenamefont {Lettner}, \citenamefont
  {Volz}, \citenamefont {Dietze}, \citenamefont {Garcia-Ripoll}, \citenamefont
  {Cirac}, \citenamefont {Rempe},\ and\ \citenamefont
  {D{\"u}rr}}]{Syassen2008}%
  \BibitemOpen
  \bibfield  {author} {\bibinfo {author} {\bibfnamefont {N.}~\bibnamefont
  {Syassen}}, \bibinfo {author} {\bibfnamefont {D.}~\bibnamefont {Bauer}},
  \bibinfo {author} {\bibfnamefont {M.}~\bibnamefont {Lettner}}, \bibinfo
  {author} {\bibfnamefont {T.}~\bibnamefont {Volz}}, \bibinfo {author}
  {\bibfnamefont {D.}~\bibnamefont {Dietze}}, \bibinfo {author} {\bibfnamefont
  {J.}~\bibnamefont {Garcia-Ripoll}}, \bibinfo {author} {\bibfnamefont
  {J.}~\bibnamefont {Cirac}}, \bibinfo {author} {\bibfnamefont
  {G.}~\bibnamefont {Rempe}}, \ and\ \bibinfo {author} {\bibfnamefont
  {S.}~\bibnamefont {D{\"u}rr}},\ }\href@noop {} {\bibfield  {journal}
  {\bibinfo  {journal} {Science}\ }\textbf {\bibinfo {volume} {320}},\ \bibinfo
  {pages} {1329} (\bibinfo {year} {2008})}\BibitemShut {NoStop}%
\bibitem [{\citenamefont {D{\"u}rr}\ \emph {et~al.}(2009)\citenamefont
  {D{\"u}rr}, \citenamefont {Garcia-Ripoll}, \citenamefont {Syassen},
  \citenamefont {Bauer}, \citenamefont {Lettner}, \citenamefont {Cirac},\ and\
  \citenamefont {Rempe}}]{Durr2009}%
  \BibitemOpen
  \bibfield  {author} {\bibinfo {author} {\bibfnamefont {S.}~\bibnamefont
  {D{\"u}rr}}, \bibinfo {author} {\bibfnamefont {J.}~\bibnamefont
  {Garcia-Ripoll}}, \bibinfo {author} {\bibfnamefont {N.}~\bibnamefont
  {Syassen}}, \bibinfo {author} {\bibfnamefont {D.}~\bibnamefont {Bauer}},
  \bibinfo {author} {\bibfnamefont {M.}~\bibnamefont {Lettner}}, \bibinfo
  {author} {\bibfnamefont {J.}~\bibnamefont {Cirac}}, \ and\ \bibinfo {author}
  {\bibfnamefont {G.}~\bibnamefont {Rempe}},\ }\href@noop {} {\bibfield
  {journal} {\bibinfo  {journal} {Physical Review A}\ }\textbf {\bibinfo
  {volume} {79}},\ \bibinfo {pages} {023614} (\bibinfo {year}
  {2009})}\BibitemShut {NoStop}%
\bibitem [{\citenamefont {Zhu}\ \emph {et~al.}(2014)\citenamefont {Zhu},
  \citenamefont {Gadway}, \citenamefont {Foss-Feig}, \citenamefont
  {Schachenmayer}, \citenamefont {Wall}, \citenamefont {Hazzard}, \citenamefont
  {Yan}, \citenamefont {Moses}, \citenamefont {Covey}, \citenamefont {Jin},
  \citenamefont {Ye}, \citenamefont {Holland},\ and\ \citenamefont
  {Rey}}]{Zhu2014}%
  \BibitemOpen
  \bibfield  {author} {\bibinfo {author} {\bibfnamefont {B.}~\bibnamefont
  {Zhu}}, \bibinfo {author} {\bibfnamefont {B.}~\bibnamefont {Gadway}},
  \bibinfo {author} {\bibfnamefont {M.}~\bibnamefont {Foss-Feig}}, \bibinfo
  {author} {\bibfnamefont {J.}~\bibnamefont {Schachenmayer}}, \bibinfo {author}
  {\bibfnamefont {M.}~\bibnamefont {Wall}}, \bibinfo {author} {\bibfnamefont
  {K.}~\bibnamefont {Hazzard}}, \bibinfo {author} {\bibfnamefont
  {B.}~\bibnamefont {Yan}}, \bibinfo {author} {\bibfnamefont {S.}~\bibnamefont
  {Moses}}, \bibinfo {author} {\bibfnamefont {J.}~\bibnamefont {Covey}},
  \bibinfo {author} {\bibfnamefont {D.}~\bibnamefont {Jin}}, \bibinfo {author}
  {\bibfnamefont {J.}~\bibnamefont {Ye}}, \bibinfo {author} {\bibfnamefont
  {M.}~\bibnamefont {Holland}}, \ and\ \bibinfo {author} {\bibfnamefont
  {A.}~\bibnamefont {Rey}},\ }\href@noop {} {\bibfield  {journal} {\bibinfo
  {journal} {Phys. Rev. Lett.}\ }\textbf {\bibinfo {volume} {112}},\ \bibinfo
  {pages} {070404} (\bibinfo {year} {2014})}\BibitemShut {NoStop}%
\bibitem [{\citenamefont {Bush}\ \emph {et~al.}(1998)\citenamefont {Bush},
  \citenamefont {Englert}, \citenamefont {Rza\ifmmode \mbox{\c{}}\else
  \c{}\fi{}\ifmmode~\dot{z}\else \.{z}\fi{}ewski},\ and\ \citenamefont
  {Wilkens}}]{Bush}%
  \BibitemOpen
  \bibfield  {author} {\bibinfo {author} {\bibfnamefont {T.}~\bibnamefont
  {Bush}}, \bibinfo {author} {\bibfnamefont {B.}~\bibnamefont {Englert}},
  \bibinfo {author} {\bibfnamefont {K.}~\bibnamefont {Rza\ifmmode
  \mbox{\c{}}\else \c{}\fi{}\ifmmode~\dot{z}\else \.{z}\fi{}ewski}}, \ and\
  \bibinfo {author} {\bibfnamefont {M.}~\bibnamefont {Wilkens}},\ }\href@noop
  {} {\bibfield  {journal} {\bibinfo  {journal} {Found. Phys.}\ }\textbf
  {\bibinfo {volume} {28}},\ \bibinfo {pages} {549} (\bibinfo {year}
  {1998})}\BibitemShut {NoStop}%
\bibitem [{\citenamefont {Tiesinga}\ \emph {et~al.}(2000)\citenamefont
  {Tiesinga}, \citenamefont {Williams}, \citenamefont {Mies},\ and\
  \citenamefont {Julienne}}]{Tiesinga2000}%
  \BibitemOpen
  \bibfield  {author} {\bibinfo {author} {\bibfnamefont {E.}~\bibnamefont
  {Tiesinga}}, \bibinfo {author} {\bibfnamefont {C.~J.}\ \bibnamefont
  {Williams}}, \bibinfo {author} {\bibfnamefont {F.~H.}\ \bibnamefont {Mies}},
  \ and\ \bibinfo {author} {\bibfnamefont {P.~S.}\ \bibnamefont {Julienne}},\
  }\href@noop {} {\bibfield  {journal} {\bibinfo  {journal} {Phys. Rev. A}\
  }\textbf {\bibinfo {volume} {61}},\ \bibinfo {pages} {063416} (\bibinfo
  {year} {2000})}\BibitemShut {NoStop}%
\bibitem [{\citenamefont {Bolda}\ \emph {et~al.}(2002)\citenamefont {Bolda},
  \citenamefont {Tiesinga},\ and\ \citenamefont {Julienne}}]{Bolda2002}%
  \BibitemOpen
  \bibfield  {author} {\bibinfo {author} {\bibfnamefont {E.~L.}\ \bibnamefont
  {Bolda}}, \bibinfo {author} {\bibfnamefont {E.}~\bibnamefont {Tiesinga}}, \
  and\ \bibinfo {author} {\bibfnamefont {P.~S.}\ \bibnamefont {Julienne}},\
  }\href@noop {} {\bibfield  {journal} {\bibinfo  {journal} {Phys. Rev. A}\
  }\textbf {\bibinfo {volume} {66}},\ \bibinfo {pages} {013403} (\bibinfo
  {year} {2002})}\BibitemShut {NoStop}%
\bibitem [{\citenamefont {Blume}\ and\ \citenamefont
  {Greene}(2002)}]{Blume2002}%
  \BibitemOpen
  \bibfield  {author} {\bibinfo {author} {\bibfnamefont {D.}~\bibnamefont
  {Blume}}\ and\ \bibinfo {author} {\bibfnamefont {C.~H.}\ \bibnamefont
  {Greene}},\ }\href@noop {} {\bibfield  {journal} {\bibinfo  {journal} {Phys.
  Rev. A}\ }\textbf {\bibinfo {volume} {65}},\ \bibinfo {pages} {043613}
  (\bibinfo {year} {2002})}\BibitemShut {NoStop}%
\bibitem [{\citenamefont {Bolda}\ \emph {et~al.}(2003)\citenamefont {Bolda},
  \citenamefont {Tiesinga},\ and\ \citenamefont {Julienne}}]{Bolda2003}%
  \BibitemOpen
  \bibfield  {author} {\bibinfo {author} {\bibfnamefont {E.~L.}\ \bibnamefont
  {Bolda}}, \bibinfo {author} {\bibfnamefont {E.}~\bibnamefont {Tiesinga}}, \
  and\ \bibinfo {author} {\bibfnamefont {P.~S.}\ \bibnamefont {Julienne}},\
  }\href@noop {} {\bibfield  {journal} {\bibinfo  {journal} {Phys. Rev. A}\
  }\textbf {\bibinfo {volume} {68}},\ \bibinfo {pages} {032702} (\bibinfo
  {year} {2003})}\BibitemShut {NoStop}%
\bibitem [{\citenamefont {Idziaszek}\ and\ \citenamefont
  {Calarco}(2005)}]{Idziaszek2005}%
  \BibitemOpen
  \bibfield  {author} {\bibinfo {author} {\bibfnamefont {Z.}~\bibnamefont
  {Idziaszek}}\ and\ \bibinfo {author} {\bibfnamefont {T.}~\bibnamefont
  {Calarco}},\ }\href@noop {} {\bibfield  {journal} {\bibinfo  {journal} {Phys.
  Rev. A}\ }\textbf {\bibinfo {volume} {71}},\ \bibinfo {pages} {050701}
  (\bibinfo {year} {2005})}\BibitemShut {NoStop}%
\bibitem [{\citenamefont {Idziaszek}\ and\ \citenamefont
  {Calarco}(2006)}]{Idziaszek2006}%
  \BibitemOpen
  \bibfield  {author} {\bibinfo {author} {\bibfnamefont {Z.}~\bibnamefont
  {Idziaszek}}\ and\ \bibinfo {author} {\bibfnamefont {T.}~\bibnamefont
  {Calarco}},\ }\href@noop {} {\bibfield  {journal} {\bibinfo  {journal} {Phys.
  Rev. A}\ }\textbf {\bibinfo {volume} {74}},\ \bibinfo {pages} {022712}
  (\bibinfo {year} {2006})}\BibitemShut {NoStop}%
\bibitem [{\citenamefont {Naidon}\ \emph {et~al.}(2007)\citenamefont {Naidon},
  \citenamefont {Tiesinga}, \citenamefont {Mitchell},\ and\ \citenamefont
  {Julienne}}]{Naidon2007}%
  \BibitemOpen
  \bibfield  {author} {\bibinfo {author} {\bibfnamefont {P.}~\bibnamefont
  {Naidon}}, \bibinfo {author} {\bibfnamefont {E.}~\bibnamefont {Tiesinga}},
  \bibinfo {author} {\bibfnamefont {W.~F.}\ \bibnamefont {Mitchell}}, \ and\
  \bibinfo {author} {\bibfnamefont {P.~S.}\ \bibnamefont {Julienne}},\
  }\href@noop {} {\bibfield  {journal} {\bibinfo  {journal} {New Journal of
  Physics}\ }\textbf {\bibinfo {volume} {9}},\ \bibinfo {pages} {19} (\bibinfo
  {year} {2007})}\BibitemShut {NoStop}%
\bibitem [{\citenamefont {Yurovsky}\ and\ \citenamefont
  {Band}(2007)}]{Yurovsky2007}%
  \BibitemOpen
  \bibfield  {author} {\bibinfo {author} {\bibfnamefont {V.~A.}\ \bibnamefont
  {Yurovsky}}\ and\ \bibinfo {author} {\bibfnamefont {Y.~B.}\ \bibnamefont
  {Band}},\ }\href@noop {} {\bibfield  {journal} {\bibinfo  {journal} {Phys.
  Rev. A}\ }\textbf {\bibinfo {volume} {75}},\ \bibinfo {pages} {012717}
  (\bibinfo {year} {2007})}\BibitemShut {NoStop}%
\bibitem [{\citenamefont {Idziaszek}(2009)}]{Idziaszek2009}%
  \BibitemOpen
  \bibfield  {author} {\bibinfo {author} {\bibfnamefont {Z.}~\bibnamefont
  {Idziaszek}},\ }\href@noop {} {\bibfield  {journal} {\bibinfo  {journal}
  {Phys. Rev. A}\ }\textbf {\bibinfo {volume} {79}},\ \bibinfo {pages} {062701}
  (\bibinfo {year} {2009})}\BibitemShut {NoStop}%
\bibitem [{\citenamefont {Yurovsky}\ \emph {et~al.}(2008)\citenamefont
  {Yurovsky}, \citenamefont {Olshanii},\ and\ \citenamefont
  {Weiss}}]{Yurovsky2008}%
  \BibitemOpen
  \bibfield  {author} {\bibinfo {author} {\bibfnamefont {V.~A.}\ \bibnamefont
  {Yurovsky}}, \bibinfo {author} {\bibfnamefont {M.}~\bibnamefont {Olshanii}},
  \ and\ \bibinfo {author} {\bibfnamefont {D.~S.}\ \bibnamefont {Weiss}},\
  }\href@noop {} {\ \bibinfo {series} {Advances In Atomic, Molecular, and
  Optical Physics},\ \textbf {\bibinfo {volume} {55}},\ \bibinfo {pages} {61 }
  (\bibinfo {year} {2008})}\BibitemShut {NoStop}%
\bibitem [{\citenamefont {Olshanii}(1998)}]{Olshanii}%
  \BibitemOpen
  \bibfield  {author} {\bibinfo {author} {\bibfnamefont {M.}~\bibnamefont
  {Olshanii}},\ }\href@noop {} {\bibfield  {journal} {\bibinfo  {journal}
  {Phys. Rev. Lett.}\ }\textbf {\bibinfo {volume} {81}},\ \bibinfo {pages}
  {938} (\bibinfo {year} {1998})}\BibitemShut {NoStop}%
\bibitem [{\citenamefont {Petrov}\ \emph {et~al.}(2000)\citenamefont {Petrov},
  \citenamefont {Holzmann},\ and\ \citenamefont {Shlyapnikov}}]{Petrov2000}%
  \BibitemOpen
  \bibfield  {author} {\bibinfo {author} {\bibfnamefont {D.~S.}\ \bibnamefont
  {Petrov}}, \bibinfo {author} {\bibfnamefont {M.}~\bibnamefont {Holzmann}}, \
  and\ \bibinfo {author} {\bibfnamefont {G.~V.}\ \bibnamefont {Shlyapnikov}},\
  }\href@noop {} {\bibfield  {journal} {\bibinfo  {journal} {Phys. Rev. Lett.}\
  }\textbf {\bibinfo {volume} {84}},\ \bibinfo {pages} {2551} (\bibinfo {year}
  {2000})}\BibitemShut {NoStop}%
\bibitem [{\citenamefont {Petrov}\ and\ \citenamefont
  {Shlyapnikov}(2001)}]{Petrov2001}%
  \BibitemOpen
  \bibfield  {author} {\bibinfo {author} {\bibfnamefont {D.}~\bibnamefont
  {Petrov}}\ and\ \bibinfo {author} {\bibfnamefont {G.}~\bibnamefont
  {Shlyapnikov}},\ }\href@noop {} {\bibfield  {journal} {\bibinfo  {journal}
  {Physical Review A}\ }\textbf {\bibinfo {volume} {64}},\ \bibinfo {pages}
  {012706} (\bibinfo {year} {2001})}\BibitemShut {NoStop}%
\bibitem [{\citenamefont {Granger}\ and\ \citenamefont
  {Blume}(2004)}]{Granger2004}%
  \BibitemOpen
  \bibfield  {author} {\bibinfo {author} {\bibfnamefont {B.~E.}\ \bibnamefont
  {Granger}}\ and\ \bibinfo {author} {\bibfnamefont {D.}~\bibnamefont
  {Blume}},\ }\href@noop {} {\bibfield  {journal} {\bibinfo  {journal}
  {Physical review letters}\ }\textbf {\bibinfo {volume} {92}},\ \bibinfo
  {pages} {133202} (\bibinfo {year} {2004})}\BibitemShut {NoStop}%
\bibitem [{\citenamefont {Kanjilal}\ and\ \citenamefont
  {Blume}(2004)}]{Kanjilal2004}%
  \BibitemOpen
  \bibfield  {author} {\bibinfo {author} {\bibfnamefont {K.}~\bibnamefont
  {Kanjilal}}\ and\ \bibinfo {author} {\bibfnamefont {D.}~\bibnamefont
  {Blume}},\ }\href@noop {} {\bibfield  {journal} {\bibinfo  {journal} {Phys.
  Rev. A}\ }\textbf {\bibinfo {volume} {70}},\ \bibinfo {pages} {042709}
  (\bibinfo {year} {2004})}\BibitemShut {NoStop}%
\bibitem [{\citenamefont {Melezhik}\ and\ \citenamefont
  {Schmelcher}(2009)}]{Melezhik2009}%
  \BibitemOpen
  \bibfield  {author} {\bibinfo {author} {\bibfnamefont {V.~S.}\ \bibnamefont
  {Melezhik}}\ and\ \bibinfo {author} {\bibfnamefont {P.}~\bibnamefont
  {Schmelcher}},\ }\href@noop {} {\bibfield  {journal} {\bibinfo  {journal}
  {New Journal of Physics}\ }\textbf {\bibinfo {volume} {11}},\ \bibinfo
  {pages} {073031} (\bibinfo {year} {2009})}\BibitemShut {NoStop}%
\bibitem [{\citenamefont {Peng}\ \emph {et~al.}(2010)\citenamefont {Peng},
  \citenamefont {Bohloul}, \citenamefont {Liu}, \citenamefont {Hu},\ and\
  \citenamefont {Drummond}}]{Drummond2010}%
  \BibitemOpen
  \bibfield  {author} {\bibinfo {author} {\bibfnamefont {S.-G.}\ \bibnamefont
  {Peng}}, \bibinfo {author} {\bibfnamefont {S.~S.}\ \bibnamefont {Bohloul}},
  \bibinfo {author} {\bibfnamefont {X.-J.}\ \bibnamefont {Liu}}, \bibinfo
  {author} {\bibfnamefont {H.}~\bibnamefont {Hu}}, \ and\ \bibinfo {author}
  {\bibfnamefont {P.~D.}\ \bibnamefont {Drummond}},\ }\href@noop {} {\bibfield
  {journal} {\bibinfo  {journal} {Phys. Rev. A}\ }\textbf {\bibinfo {volume}
  {82}},\ \bibinfo {pages} {063633} (\bibinfo {year} {2010})}\BibitemShut
  {NoStop}%
\bibitem [{\citenamefont {Peng}\ \emph {et~al.}(2011)\citenamefont {Peng},
  \citenamefont {Hu}, \citenamefont {Liu},\ and\ \citenamefont
  {Drummond}}]{Drummond2011}%
  \BibitemOpen
  \bibfield  {author} {\bibinfo {author} {\bibfnamefont {S.-G.}\ \bibnamefont
  {Peng}}, \bibinfo {author} {\bibfnamefont {H.}~\bibnamefont {Hu}}, \bibinfo
  {author} {\bibfnamefont {X.-J.}\ \bibnamefont {Liu}}, \ and\ \bibinfo
  {author} {\bibfnamefont {P.~D.}\ \bibnamefont {Drummond}},\ }\href@noop {}
  {\bibfield  {journal} {\bibinfo  {journal} {Phys. Rev. A}\ }\textbf {\bibinfo
  {volume} {84}},\ \bibinfo {pages} {043619} (\bibinfo {year}
  {2011})}\BibitemShut {NoStop}%
\bibitem [{\citenamefont {Sala}\ \emph {et~al.}(2012)\citenamefont {Sala},
  \citenamefont {Schneider},\ and\ \citenamefont {Saenz}}]{Sala2012}%
  \BibitemOpen
  \bibfield  {author} {\bibinfo {author} {\bibfnamefont {S.}~\bibnamefont
  {Sala}}, \bibinfo {author} {\bibfnamefont {P.-I.}\ \bibnamefont {Schneider}},
  \ and\ \bibinfo {author} {\bibfnamefont {A.}~\bibnamefont {Saenz}},\
  }\href@noop {} {\bibfield  {journal} {\bibinfo  {journal} {Physical review
  letters}\ }\textbf {\bibinfo {volume} {109}},\ \bibinfo {pages} {073201}
  (\bibinfo {year} {2012})}\BibitemShut {NoStop}%
\bibitem [{\citenamefont {Moritz}\ \emph {et~al.}(2005)\citenamefont {Moritz},
  \citenamefont {{S}t\"oferle}, \citenamefont {{G}\"unter}, \citenamefont
  {{K}\"ohl},\ and\ \citenamefont {{E}sslinger}}]{Moritz2005}%
  \BibitemOpen
  \bibfield  {author} {\bibinfo {author} {\bibfnamefont {H.}~\bibnamefont
  {Moritz}}, \bibinfo {author} {\bibfnamefont {T.}~\bibnamefont
  {{S}t\"oferle}}, \bibinfo {author} {\bibfnamefont {K.}~\bibnamefont
  {{G}\"unter}}, \bibinfo {author} {\bibfnamefont {M.}~\bibnamefont
  {{K}\"ohl}}, \ and\ \bibinfo {author} {\bibfnamefont {T.}~\bibnamefont
  {{E}sslinger}},\ }\href@noop {} {\bibfield  {journal} {\bibinfo  {journal}
  {Phys. {R}ev. {L}ett.}\ }\textbf {\bibinfo {volume} {94}},\ \bibinfo {pages}
  {210401} (\bibinfo {year} {2005})}\BibitemShut {NoStop}%
\bibitem [{\citenamefont {Haller}\ \emph {et~al.}(2010)\citenamefont {Haller},
  \citenamefont {Mark}, \citenamefont {Hart}, \citenamefont {Danzl},
  \citenamefont {Reichs{\"o}llner}, \citenamefont {Melezhik}, \citenamefont
  {Schmelcher},\ and\ \citenamefont {N{\"a}gerl}}]{Haller2010}%
  \BibitemOpen
  \bibfield  {author} {\bibinfo {author} {\bibfnamefont {E.}~\bibnamefont
  {Haller}}, \bibinfo {author} {\bibfnamefont {M.~J.}\ \bibnamefont {Mark}},
  \bibinfo {author} {\bibfnamefont {R.}~\bibnamefont {Hart}}, \bibinfo {author}
  {\bibfnamefont {J.~G.}\ \bibnamefont {Danzl}}, \bibinfo {author}
  {\bibfnamefont {L.}~\bibnamefont {Reichs{\"o}llner}}, \bibinfo {author}
  {\bibfnamefont {V.}~\bibnamefont {Melezhik}}, \bibinfo {author}
  {\bibfnamefont {P.}~\bibnamefont {Schmelcher}}, \ and\ \bibinfo {author}
  {\bibfnamefont {H.-C.}\ \bibnamefont {N{\"a}gerl}},\ }\href@noop {}
  {\bibfield  {journal} {\bibinfo  {journal} {Physical review letters}\
  }\textbf {\bibinfo {volume} {104}},\ \bibinfo {pages} {153203} (\bibinfo
  {year} {2010})}\BibitemShut {NoStop}%
\bibitem [{\citenamefont {Sala}\ \emph {et~al.}(2013)\citenamefont {Sala},
  \citenamefont {Z\"urn}, \citenamefont {Lompe}, \citenamefont {Wenz},
  \citenamefont {Murmann}, \citenamefont {Serwane}, \citenamefont {Jochim},\
  and\ \citenamefont {Saenz}}]{Sala2013}%
  \BibitemOpen
  \bibfield  {author} {\bibinfo {author} {\bibfnamefont {S.}~\bibnamefont
  {Sala}}, \bibinfo {author} {\bibfnamefont {G.}~\bibnamefont {Z\"urn}},
  \bibinfo {author} {\bibfnamefont {T.}~\bibnamefont {Lompe}}, \bibinfo
  {author} {\bibfnamefont {A.~N.}\ \bibnamefont {Wenz}}, \bibinfo {author}
  {\bibfnamefont {S.}~\bibnamefont {Murmann}}, \bibinfo {author} {\bibfnamefont
  {F.}~\bibnamefont {Serwane}}, \bibinfo {author} {\bibfnamefont
  {S.}~\bibnamefont {Jochim}}, \ and\ \bibinfo {author} {\bibfnamefont
  {A.}~\bibnamefont {Saenz}},\ }\href@noop {} {\bibfield  {journal} {\bibinfo
  {journal} {Phys. Rev. Lett.}\ }\textbf {\bibinfo {volume} {110}},\ \bibinfo
  {pages} {203202} (\bibinfo {year} {2013})}\BibitemShut {NoStop}%
\bibitem [{\citenamefont {Naidon}\ and\ \citenamefont
  {Julienne}(2006)}]{Naidon2006}%
  \BibitemOpen
  \bibfield  {author} {\bibinfo {author} {\bibfnamefont {P.}~\bibnamefont
  {Naidon}}\ and\ \bibinfo {author} {\bibfnamefont {P.~S.}\ \bibnamefont
  {Julienne}},\ }\href@noop {} {\bibfield  {journal} {\bibinfo  {journal}
  {{Phys. Rev. A}}\ }\textbf {\bibinfo {volume} {{74}}},\ \bibinfo {pages}
  {{062713}} (\bibinfo {year} {{2006}})}\BibitemShut {NoStop}%
\bibitem [{\citenamefont {Jachymski}\ \emph {et~al.}(2013)\citenamefont
  {Jachymski}, \citenamefont {Krych}, \citenamefont {Julienne},\ and\
  \citenamefont {Idziaszek}}]{Jachymski2013}%
  \BibitemOpen
  \bibfield  {author} {\bibinfo {author} {\bibfnamefont {K.}~\bibnamefont
  {Jachymski}}, \bibinfo {author} {\bibfnamefont {M.}~\bibnamefont {Krych}},
  \bibinfo {author} {\bibfnamefont {P.~S.}\ \bibnamefont {Julienne}}, \ and\
  \bibinfo {author} {\bibfnamefont {Z.}~\bibnamefont {Idziaszek}},\ }\href@noop
  {} {\bibfield  {journal} {\bibinfo  {journal} {Phys. Rev. Lett.}\ }\textbf
  {\bibinfo {volume} {110}},\ \bibinfo {pages} {213202} (\bibinfo {year}
  {2013})}\BibitemShut {NoStop}%
\bibitem [{\citenamefont {Girardeau}\ \emph {et~al.}(2004)\citenamefont
  {Girardeau}, \citenamefont {Nguyen},\ and\ \citenamefont
  {Olshanii}}]{Girardeau2004}%
  \BibitemOpen
  \bibfield  {author} {\bibinfo {author} {\bibfnamefont {M.}~\bibnamefont
  {Girardeau}}, \bibinfo {author} {\bibfnamefont {H.}~\bibnamefont {Nguyen}}, \
  and\ \bibinfo {author} {\bibfnamefont {M.}~\bibnamefont {Olshanii}},\
  }\href@noop {} {\bibfield  {journal} {\bibinfo  {journal} {Optics
  Communications}\ }\textbf {\bibinfo {volume} {243}},\ \bibinfo {pages} {3}
  (\bibinfo {year} {2004})}\BibitemShut {NoStop}%
\end{thebibliography}%

\end{document}